\documentclass{article}
\usepackage{codefuse_tech_report}
\usepackage[colorlinks = true,
            linkcolor = blue,
            urlcolor  = blue,
            citecolor = blue,
            anchorcolor = blue]{hyperref}   
\usepackage{microtype}
\usepackage{xurl}
\usepackage{booktabs}
\usepackage{enumitem}
\usepackage{multicol}
\usepackage{CJKutf8}
\usepackage{colortbl}
\usepackage{xcolor}
\usepackage{color}
\usepackage{siunitx}
\usepackage{floatflt}
\usepackage{graphicx}
\usepackage{wrapfig}
\usepackage{xspace}
\usepackage{authblk}
\usepackage{lipsum}
\usepackage{algorithm}
\usepackage{algorithmicx}
\usepackage{algpseudocode}
\usepackage{subfigure}
\usepackage{multirow}
\usepackage{pifont}  
\usepackage{listings, multicol}

\algnewcommand{\LeftComment}[1]{\Statex \(\triangleright\) #1}

\usepackage{array}
\usepackage{amsmath}
\usepackage{amssymb}
\usepackage{mathtools}
\usepackage{amsthm}

\usepackage[capitalize,noabbrev]{cleveref}
\usepackage{adjustbox} 

\theoremstyle{plain}

\theoremstyle{definition}

\theoremstyle{remark}

\newcommand{\ie}{\textit{i.e.}}
\newcommand{\eg}{\textit{e.g.}}

\colmfinalcopy

\usepackage[utf8]{inputenc}
\usepackage[T1]{fontenc}
\usepackage{caption} 
\usepackage{adjustbox} 
\usepackage{fontawesome5}

\usepackage{tcolorbox}
\tcbuselibrary{skins,breakable}

\tcbuselibrary{skins}

\usepackage{color}
\usepackage{xcolor}
\usepackage{soul} 

\definecolor{tred}{RGB}{251, 130, 132}
\definecolor{torange}{RGB}{247, 162, 116}
\definecolor{tyellow}{RGB}{251, 218, 140}
\definecolor{tgreen}{RGB}{127, 204, 181}
\definecolor{tblue}{RGB}{89, 177, 215}
\definecolor{insightblue}{RGB}{162, 210, 255}
\definecolor{questionred}{RGB}{255, 175, 204}

\newcommand{\lstbg}[3][0pt]{{\fboxsep#1\colorbox{#2}{\strut #3}}}
\definecolor{codegreen}{rgb}{0,0.6,0}
\lstdefinelanguage{diff}{
	frame=shadowbox,
	basicstyle=\ttfamily\scriptsize\bfseries,
        breaklines=true,
	morecomment=[f][\color{red}]{---}, 
	morecomment=[f][\color{codegreen}]{+++},
        morecomment=[f][\color{orange}]{<<<<<<<},
        morecomment=[f][\color{orange}]{=======},
        morecomment=[f][\color{orange}]{>>>>>>>},
	morecomment=[f][\lstbg{red!20}]{-\ },
	morecomment=[f][\lstbg{green!20}]{+\ },
	morecomment=[f][\color{blue}]{@@},
}

\usepackage[framemethod=TikZ]{mdframed}
\usepackage{lipsum}
\mdfsetup{skipabove=7pt,skipbelow=0pt}
\definecolor{darkcyan}{HTML}{0091A4}  
\definecolor{brightcyan}{HTML}{dcf0f2}
\newcounter{conclusioncounter}
\mdfdefinestyle{ccl}{
    linecolor=darkcyan, 
    outerlinewidth=3pt,
    leftline=true, 
    rightline=false, 
    topline=false, 
    bottomline=false, 
    innertopmargin=10pt, 
    innerbottommargin=10pt, 
    innerrightmargin=13pt, 
    innerleftmargin=10pt, 
    backgroundcolor=brightcyan 
}
\newcommand{\ccl}{\refstepcounter{conclusioncounter}\textbf{Conclusion:}\xspace}

\title{CodeFuse-CommitEval: Towards Benchmarking LLM's Power on Commit Message and Code Change Inconsistency Detection}

\newcommand{\sys}{\textsc{CodeFuse-CommitEval}\xspace}
\newcommand{\etal}{{\em et al.}\xspace}

\definecolor{mygreen}{HTML}{02818a}

\author{%
Qingyu Zhang$^{1,2}$
~~Puzhuo Liu$^{1,3}$
~~Peng Di$^{1,4}$
~~Chenxiong Qian$^{2}$
\\

\vspace{10pt}
$^1$Ant Group\ \ \ $^2$The University of Hong Kong\\ $^3$Tsinghua University\ \ \ $^4$The University of New South Wales\\
\vspace{10pt}
\hspace{-10pt}\faGithub ~\url{https://github.com/codefuse-ai/CodeFuse-CommitEval}\\
}

\colmfinalcopy 

\begin{document}

\maketitle

\begin{abstract}
Version control relies on commit messages to convey the rationale for code changes, but these messages are often low quality and, more critically, inconsistent with their diffs—known as message-code inconsistency (MCI). 
MCIs mislead reviewers, hinder maintenance, contaminate research datasets, and may obscure security patches. Yet, no dedicated benchmark exists to evaluate models for MCI detection. We introduce CODEFUSE-COMMITEVAL, the first benchmark designed for MCI detection using large language models. 
Built on the ApacheCM dataset for diversity and quality, we generate seven types of inconsistent messages through rule-guided mutations of originally consistent commits and apply two-fold validation to verify both positive and negative samples. Using this labeled dataset of message-diff pairs, we evaluate six state-of-the-art open-source LLMs under a vanilla setting and with three augmentation strategies: few-shot prompting, chain-of-thought, and extended context. Results show models detect inconsistent commits more reliably than consistent ones (average Recall 85.95\%, Precision 80.28\%, Specificity 63.8\%); gpt-oss-20B performs best overall but uses over twice the tokens of others. Augmentation effects vary: adjacent context helps larger models but adds noise for smaller ones; few-shot improves accuracy and reduces token use, yet increases universally incorrect predictions; chain-of-thought boosts precision and specificity at the cost of recall and higher token consumption. Type-wise analysis reveals higher detectability for component, file-path, and operation inconsistencies, but lower accuracy and higher token cost for intent-level "purpose" inconsistencies. CODEFUSE-COMMITEVAL provides a rigorous foundation for measuring, comparing, and advancing MCI detection, highlighting the need for richer context and balanced data to capture high-level semantic gaps.
\end{abstract}


\section{Introduction}

%
Version control systems~\cite{vcs} are foundational to modern software development, enabling developers to collaboratively manage, track, and maintain changes to a codebase over time.
These systems facilitate essential operations such as branching, merging, and reverting changes, while providing a robust audit trail of the project’s evolution~\cite{zolkifli2018version, koc2011survey}.
At the core of version control lies the commit, which encapsulates one or more atomic changes along with a descriptive message~\cite{chacon2014pro, commitstudy1}.
Commits are not only the basic unit of repository evolution but also the primary vehicle through which knowledge about code modifications is communicated.
Thus, the reliability of both code and its accompanying message is critical to preserving the integrity of the development history.

Despite this central role, the quality and consistency of commit messages are frequently overlooked.
Well-structured messages are vital for communication, program comprehension, and long-term maintenance, particularly in large or rapidly evolving projects where future developers depend on accurate records of past changes~\cite{commitstudy1}.
However, empirical studies show that commit message quality often deteriorates over time: for example, in one study of five active open-source projects, about 44\% of commit messages lacked sufficient detail~\cite{commitstudy2}; in another study of more than 23,000 SourceForge projects, 14\% of commit messages were entirely blank~\cite{li2025optimization}.
These findings highlight a persistent gap between the importance of commit messages and the reality of their practice.

Beyond issues of quality, a more severe problem arises when commit messages are inconsistent with the underlying code changes, a phenomenon we term message-code inconsistency (MCI).
Such inconsistencies undermine the reliability of commit histories, misleading reviewers, confusing maintainers, and ultimately eroding trust in the version control system itself.
In addition, some tangling commits, which bundle unrelated changes into a single commit, further complicate code review and debugging and raise the occurrence ratio of inconsistency.
By mixing multiple purposes, they obscure the intent behind changes, hinder program comprehension, and reduce the separation of concerns~\cite{li2022utango,kirinuki2014hey,dias2015untangling,muylaert2018untangling,herbold2022fine}.
Prior work has shown that misleading or abnormal commit messages correlate with build failures~\cite{santos2016judging} and impair developers’ ability to understand code, thereby increasing the likelihood of future defects~\cite{commitstudy1}.
More critically, these inconsistencies can propagate into datasets constructed from commit histories, thereby influencing the downstream evaluation of AI models.
For example, a prior study~\cite{herzig2013impact} found that inconsistent commits introduce substantial noise into change datasets, affecting up to 20\% of bug fixes involving multiple changes.
This noise can cause at least 16.6\% of source files to be incorrectly linked to bug reports when training models for bug prediction, directly undermining the validity of empirical results.
From a security perspective, the consequences are even more alarming: incomplete or inconsistent commit messages can obscure records of security patches~\cite{reis2023secomlint}, leading to vulnerabilities being missed or misclassified and thereby compromising both software reliability and compliance with security standards.
These findings highlight the urgent need to address message-code inconsistency during the commit review process.

Despite the clear importance of commit-message consistency, research on automatically detecting inconsistencies between commit messages and code remains limited.
Most existing work focuses on related tasks, such as detecting mismatches between code and documentation or between code and comments~\cite{lee2025metamon,zhang2024detecting,rong2024code,xu2024code,zhu2025comment}.
Yet commits differ fundamentally: while comments and documentation usually describe a single version of the code, commit messages uniquely capture how the code evolves between versions.
This difference makes existing techniques unsuitable for directly addressing MCI.

With the rapid advances of large language models (LLMs) in both natural language understanding and code reasoning, a natural opportunity arises to apply them to MCI detection.
Indeed, LLMs have already been widely studied in related areas such as commit message generation.
Benchmarks like CommitBench~\cite{schall2024commitbench} and others~\cite{wu2025empirical,jiang2017automatically,xiong2025contextual,tao2021evaluation,eliseeva2023commit,zeng2025evaluating,wu2024commit} curate pairs of code diffs and commit messages to evaluate consistency.
However, these benchmarks are designed primarily for message generation and contain almost exclusively consistent examples.
They neither annotate semantic inconsistencies nor provide sufficient positive samples for evaluating MCI detection directly.
As a result, the true ability of LLMs to recognize and flag inconsistent commit messages remains largely unmeasured. 
To this end, we propose \sys to assess the model's ability on MCI detection tasks.
To the best of our knowledge, \sys is the first benchmarking work tailored for comprehensively evaluating LLM's MCI detection ability.
\sys provides insightful findings to further related research by exploring the following three research questions.

Firstly, we aim to evaluate the fundamental performance of models on MCI detection. 
Thus, we ask \textbf{RQ1: How are the basic MCI detection abilities of various LLMs?} 
Due to the lack of a commit dataset with evenly distributed positive and negative samples, we construct \sys's dataset based on the prior commit dataset, ApacheCM~\cite{xiong2025contextual}, because of its diversity in programming languages and the high-quality commits. 
We define seven mutation rules to guide powerful LLMs in generating various types of inconsistent commit messages, based on originally consistent commit messages and the corresponding code diff. 
We then employ a two-fold forward validation to: 
1) validate the generated positive examples, and 2) validate negative examples that originally exist in the dataset. 
Each item of the verified dataset contains a commit message, the corresponding code diff, and the ground truth label indicating whether this is a consistent or inconsistent commit. 
To answer this research question, we feed the data to six state-of-the-art open-sourced models without any additional information, allowing the models to detect whether the commit is consistent or not, and compare the detection result with the ground truth label. 
The results show that all models are relatively more effective at detecting inconsistent commits, achieving an average \textit{Recall} of 85.95\% and \textit{Precision} of 80.28\%. 
Detection of consistent commits is weaker, with an average \textit{Specificity} of 63.8\%.
\textit{gpt-oss-20B} achieves the highest overall performance but incurs over twice the token usage of other models. 

Secondly, some LLM augmentation strategies have been proven effective in resolving other tasks~\cite{fewshot1,fewshot2,ctx1,ctx2,cot1,cot2}, however, the extent to which these strategies can enhance LLMs' performance on MCI detection remains unknown. 
This leads to \textbf{RQ2: Can the general LLMs augmentation strategies improve the model’s MCI detection ability?}
We leverage the aforementioned dataset to answer this research question as well.
Unlike the vanilla prompt in RQ1, which only asks the models to detect inconsistency by providing the message and the code diff, we incorporate the few-shot strategy by offering both consistent and inconsistent detection cases, and the chain-of-thought (CoT) strategy with detailed instructions specifying how a human code reviewer would detect the potential inconsistency in the commit. 
Additionally, we prepare a longer-context vanilla prompt by providing the original code diff with an extra 20 lines of code adjacent to the current diff block, to investigate the influence of additional context. 
Based on the evaluation results, we found that augmentation strategies have heterogeneous effects across inconsistency types. 
Providing adjacent code lines benefits larger models but can introduce noise for smaller ones. 
Few-shot prompting consistently improves detection accuracy and reduces token usage, yet it produces the most universally incorrect commit detection results, indicating a trade-off between stability and coverage. 
CoT prompting improves precision and specificity but at the cost of recall and substantially higher token usage. 
Overall, no single strategy uniformly dominates across all inconsistency types, highlighting that model performance varies depending on both the inconsistency pattern and the chosen augmentation method.

Finally, the type of inconsistency may also contribute to the complexity of MCI detection.
For example, if the inconsistency occurs in the message's description of changed files, it should theoretically be easy to identify by simply checking the actual changed file's name.
In contrast, if the inconsistency lies in describing the functionality of the changed code, the models must delve deeper to understand the code. 
In this case, the limited code comprehension ability of models may lead to error-prone detection results.
We propose \textbf{RQ3: Is the MCI detection ability of LLMs evenly distributed under different inconsistency patterns?} to explore the impact of different inconsistency types.
Since \sys's dataset is constructed with pre-defined mutation rules, each inconsistent commit has been labeled with its specific type.
Based on the evaluation results from previous RQs, we split the results and further classify them according to the seven inconsistency types, systematically studying the impact and correlation of the inconsistency types on the model's MCI detection ability and various augmentation strategies. 
We found that the models' MCI detection performance varies significantly across inconsistency types. 
Component, file path, and operation inconsistencies show higher detection ratios and lower token usage.
In contrast, purpose inconsistencies require deeper semantic understanding and contextual reasoning, leading to lower accuracy and substantially higher token consumption. 
These findings indicate that enhancing detection of high-level, intent-driven inconsistencies requires richer contextual grounding and more balanced datasets.

In summary, we make the following contributions:

\begin{itemize}[topsep=0pt,itemsep=0pt,parsep=0pt]
    \item We present a comprehensive pipeline for constructing MCI datasets. By combining the data synthesis capabilities of LLMs with existing commit corpora, we are able to generate a balanced dataset with both positive and negative samples, enabling rigorous evaluation of MCI detection methods. The dataset is publicly available at \url{https://figshare.com/s/21fe4ec9cb960b52bffe}.
    
    \item We introduce \sys, the first benchmark specifically designed for evaluating LLMs on the MCI detection task. Beyond measuring the baseline performance of MCI detection models, \sys also provides a platform for systematically assessing the impact of common LLM enhancement strategies on this task.
    
    \item We conduct extensive experiments on \sys with several state-of-the-art models, yielding insightful findings that highlight current limitations and point to promising directions for future research in MCI detection.
\end{itemize}
\section{Background}

\subsection{Commit Message - Code Change Inconsistency}

A Message-Code Inconsistency (MCI) occurs when the natural language description in a commit message does not accurately reflect the actual modifications in the associated code diff. 
In other words, the commit message and the code change are semantically misaligned: the message may omit, exaggerate, misrepresent, or fail to mention parts of the change, or describe intentions that differ from what the code actually does.
Given that commit messages play a critical role in software development, such inconsistencies can undermine the integrity of a project repository. 
For instance, future developers often rely on commit messages to understand why changes were made. 
If these messages mislead or omit key actions, comprehension of the system’s evolution becomes more difficult, which increases the risk of regressions. 

\textit{Inconsistent or inaccurate commit messages may mask underlying issues.} 
For example, a commit message may purport to fix a bug, but the fix may be incomplete, leaving downstream bugs undetected. 
Over time, this contributes to increased defect density within the codebase.
More severely, MCI can introduce serious security risks in real-world projects. 
One notable example occurred in 2018 with the widely-used JavaScript library \texttt{event-stream}\footnote{\url{https://github.com/dominictarr/event-stream/issues/116}}. 
Initially, the \texttt{event-stream} library was thought to be a reliable dependency for handling event streams. 
However, a malicious commit introduced a new dependency, \texttt{flatmap-stream}, that contained a hidden backdoor.
This backdoor was designed to steal Bitcoin wallet private keys by intercepting and forwarding sensitive information. 
The commit message for the update described a routine update, but the actual code diff included the addition of this harmful dependency, which was not immediately obvious to developers or maintainers.
The backdoor went unnoticed for several months, impacting projects that depended on \texttt{event-stream} without proper scrutiny of the new dependency. 
When discovered, the issue led to significant reputational damage, a costly security audit, and a lengthy process of identifying and mitigating the problem. 
This highlights how minor discrepancies between commit messages and the code diff can lead to severe security vulnerabilities and operational challenges, especially when they go unnoticed during code review.

To mitigate such problems, prominent projects enforce strict guidelines on the commit process. 
For example, when submitting patches to the Linux Kernel~\footnote{\url{https://docs.kernel.org/process/submitting-patches.html?utm_source=chatgpt.com}}, contributors are required to clearly describe both the reason for the change (“why”) and its contents (“what”) in the commit message. 
For changes that involve security patches or kernel interfaces, additional notification procedures are mandated. 
In addition, the Conventional Commits specification~\footnote{\url{https://www.conventionalcommits.org/en/v1.0.0/}} provides a lightweight, standardized format for Git commit messages. Its primary objective is to make commit messages both human-readable and machine-readable, supporting tasks such as automated versioning, CHANGELOG generation, and detection of breaking changes. The format places particular emphasis on structuring commit messages with type, optional scope, description, body, and footer fields.

With the recent advances in LLMs, especially their natural language understanding and code reasoning capabilities, there has been substantial interest in leveraging them for commit message generation.
Previous studies have shown that LLMs can produce commit messages that exceed those generated by traditional specialized models or heuristic methods~\cite{wu2025empirical}. 
However, despite such promising performance in commit generation, LLMs are known to suffer from hallucination and semantic misalignment issues: they may invent details that are not present in the code diff, misattribute changes, exaggerate or omit modifications, or otherwise produce messages that do not truthfully reflect what the code actually does~\cite{zhang2025llm,lee2025hallucination}.

\subsection{Limitations of Existing Works}

Over the past decades, researchers have struggled with detecting inconsistencies between textual descriptions (e.g., comments, documentation, and commit messages) and the corresponding source code.
Early efforts primarily relied on rule-based approaches, which could only capture narrow classes of inconsistencies~\cite{tan2012tcomment,ratol2017detecting}.
With the advent of LLMs, developers have gained more powerful tools that can reason over both natural language and code, enabling inconsistency detection in a more general and reliable manner.
Nevertheless, although prior work has made progress in detecting code–comment inconsistencies~\cite{zhu2025comment,zhang2024leveraging,zhong2025ccisolver} and code-documentation inconsistencies~\cite{lee2025metamon}, the problem of Message–Code Inconsistency (MCI) detection remains largely unexplored.
A key challenge is that commit messages fundamentally differ from comments and documentation.
Comments are typically tied to specific code segments, often clarifying the role of a variable, function, or statement in detail.
Documentation, in contrast, emphasizes higher-level abstractions of the system’s architecture and provides comprehensive descriptions of its functionality.
Both comments and documentation generally describe the semantics of a single, static version of the code.
Commit messages, however, serve a distinct role: they capture the differences between two successive versions of the code, explicitly highlighting how the system evolves across revisions.
This cross-version nature introduces unique challenges, making it difficult to directly transfer methods from related domains to the MCI setting.

Leveraging the reasoning capabilities of LLMs offers a promising direction to address this gap.
However, the first step is to establish a comprehensive understanding of LLMs’ baseline ability to reason about and detect inconsistencies between commit messages and their associated code changes.
Although several commit-related benchmarks exist~\cite{schall2024commitbench,wu2025empirical,jiang2017automatically,xiong2025contextual,tao2021evaluation,eliseeva2023commit}, these datasets are not suitable for MCI detection.
This is because they primarily focus on evaluating LLMs’ commit message generation capabilities, and thus consist almost exclusively of negative samples (i.e., consistent commit–code pairs), while lacking positive samples (i.e., inconsistent pairs).
Furthermore, LLM performance can be significantly enhanced through general augmentation strategies such as in-context learning and chain-of-thought prompting.
Evaluating the impact of these strategies on MCI detection provides not only insights into the limitations of current models but also guidance for future research directions.
\section{\sys's Pipeline}

\subsection{Overview}

\begin{figure*}[th]
	\centerline{\includegraphics[width=\linewidth]{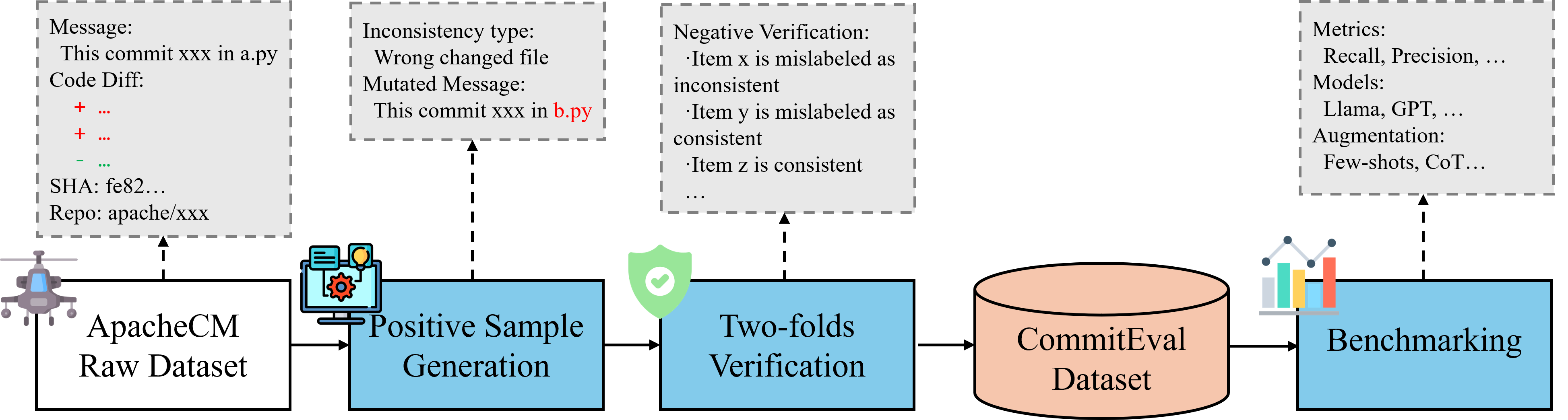}}
	\caption{\textmd{Pipeline of \sys.}}
	\label{fg:pipeline}
\end{figure*}

We illustrate the overall pipeline of \sys as shown in ~\autoref{fg:pipeline}.
\sys's dataset is constructed based on ApacheCM~\cite{xiong2025contextual}, a large-scale, multilingual commit dataset with detailed metadata and strictly filtered commit messages.
However, the original ApacheCM dataset contains only negative samples, making it unsuitable for evaluating LLMs' capabilities on MCI detection tasks.
To address this, we introduce a positive sample generation stage that leverages the powerful data synthesis abilities of LLMs. 
In this stage, a given commit message is mutated into a new message that is inconsistent with its corresponding code diff.
To ensure the validity of both positive and negative samples, we apply a two-fold verification process to filter out: 1) positive samples that are \textbf{consistent} with the actual code changes, and 2) negative samples that are \textbf{inconsistent} with the actual code changes.
\sys benchmarks selected models on this dataset, quantifying their MCI detection capabilities using rich metrics (\eg, recall and precision).
Furthermore, \sys introduces multiple model augmentation approaches to assess whether they can enhance LLMs' MCI detection performance.
The evaluation results can also be grouped according to the types of inconsistency, providing a clearer understanding of LLMs' detection performance across various inconsistent commit types.

\subsection{Positive Sample Generation}

\begin{figure*}[th]
	\centerline{\includegraphics[width=\linewidth]{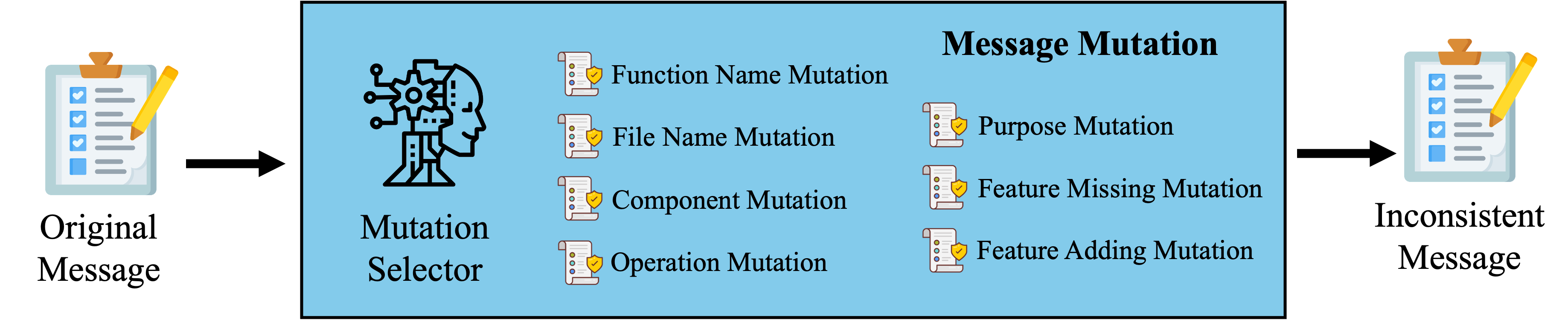}}
	\caption{\textmd{\sys's positive sample generation stage.}}
	\label{fg:sample_gen}
\end{figure*}

We illustrate the workflow for generating positive samples in~\autoref{fg:sample_gen}.
To capture the majority of MCI scenarios, we predefine seven mutation strategies:

\begin{enumerate}

\item \textbf{Function Name Mutation:} Replace mentioned function names with different but plausible names that do not actually appear in the code diff.
\item \textbf{File Name Mutation:} Replace mentioned file paths or filenames with plausible alternatives that are not among the files actually changed.
\item \textbf{Component Mutation:} Replace mentioned component, module, or system names with plausible alternatives that are not actually modified.
\item \textbf{Operation Mutation:} Alter the described operation type (\eg, from \textit{add} to \textit{remove}) so that it no longer matches the actual code changes.
\item \textbf{Purpose Mutation:} Modify the stated intent or purpose of the change while keeping the described technical action similar.
\item \textbf{Feature Missing Mutation:} Describe only a subset of the actual modifications, omitting significant changes.
\item \textbf{Feature Adding Mutation:} Describe the actual modifications and add an extra plausible, yet non-existent change.
\end{enumerate}

However, not all strategies are applicable to every commit message.
For example, if a commit message contains no function names or file paths, \textbf{Function Name Mutation} and \textbf{File Name Mutation} cannot be applied since there is no relevant information to alter.
Applying such unsuitable strategies would produce low-quality commits that fail to effectively evaluate an LLM’s capabilities.
To address this, we employ \textit{Claude Sonnet 4}\footnote{\url{https://www.anthropic.com/claude/sonnet}}, a state-of-the-art large language model, as a \textit{mutation strategy selector}.
It recommends the most suitable strategies for each commit message and finishes the mutation with specific instructions of the selected strategy's rule.
When multiple strategies are applicable, it selects the one that has been used the least so far in order to maintain the diversity of inconsistency types in our dataset as much as possible.

\subsection{Two-folds Verification}\label{sec:verification}

During the positive sample mutation stage, we observed that some mutated commit messages remained consistent with the corresponding code diffs. 
Moreover, the original messages in the dataset cannot be guaranteed to be fully aligned with the code diffs, since the filtering process in ApacheCM’s dataset construction only removes low-quality commits but does not verify the semantic consistency between commit messages and the associated code changes~\cite{xiong2025contextual}. 
For an MCI-oriented dataset, such potential noise may lead to inaccurate benchmarking results for \sys.

Given the reliable performance of LLMs in commit message generation reported in prior studies~\cite{zhang2024automatic,schall2024commitbench}, we employ \textit{Claude Sonnet 4} to generate concise and accurate commit messages from given code diffs, and to assess the semantic consistency between different versions of commit messages.
In the negative sample verification stage, the model first generates a commit message, denoted as $M_g$, based on the given code diff. 
It then determines whether $M_g$ is semantically equivalent to the original human-written commit message $M_o$ from the dataset. 
If $M_g \neq M_o$, the data point is marked as a potential error and removed from the final dataset.
Similarly, in the positive sample verification stage, the model again generates $M_g$ from the code diff.
It then checks whether $M_g$ is semantically consistent with both the original human-written message $M_o$ and the mutated message $M_m$. 
If $M_g \neq M_m$ and $M_g = M_o$, this indicates that the mutated message is indeed inconsistent with the ground truth. 
In all other cases, the corresponding positive samples are removed from the \sys dataset.
Note that $M_g$ is used solely as an intermediate result to verify the correctness of $M_o$. 
In the final dataset, for negative samples, we retain only $M_o$ in order to preserve the original semantics of the human-written content.

\subsection{Benchmarking}

In this section, we describe how \sys performs benchmarking on targeted LLMs.
We design three prompt templates to instruct the LLMs to complete MCI detection tasks, each incorporating a different general performance enhancement strategy:

\begin{enumerate}
\item \textbf{Vanilla Template:} The model is directly prompted to determine whether the given commit message and code change are consistent, along with detailed reasoning.
The complete prompt is provided in ~\autoref{apd:templates}'s ~\autoref{lst:vanilla_template}.
\item \textbf{CoT Template:} The model is instructed to follow a chain-of-thought process, simulating the reasoning steps a human code reviewer might take to identify MCI. 
The complete prompt is provided in ~\autoref{apd:templates}'s ~\autoref{lst:fewshot_template}.
\item \textbf{Few-Shot Template:} The model is provided with one positive and one negative example as references, helping it better understand the core patterns of inconsistencies. 
The complete prompt is provided in ~\autoref{apd:templates}'s ~\autoref{lst:cot_template}.
\end{enumerate}

For all templates, the commit message and code diff are provided as input to the model.
We then collect the model’s output only if it gives a clear, definitive answer (\ie, consistent or inconsistent) in the format of JSON as requested in the prompt.
Since the ground truth labels for each sample are known, we can directly evaluate the correctness of the model’s prediction.
For the vanilla template, we also experiment with providing the code diff along with either 0 or 20 lines of adjacent context around each diff block.
This is achieved by setting the \texttt{unified} parameter of the \texttt{git diff} command to 0 or 20, respectively.
For other templates, we provide them with 0-line-context code diff by default.
\section{\sys Dataset Summary}

\begin{table*}[]
\centering
\caption{\textmd{Data distributions of various types of positive and negative samples in \sys's dataset after each construction stage. The abbreviations \textit{FNM, FPM, CM, OM, PM, FMM,} and \textit{FAM} denote the mutation rules \textit{Function Name}, \textit{File Path}, \textit{Component}, \textit{Operation}, \textit{Purpose}, \textit{Feature Missing}, and \textit{Feature Adding}, respectively. The rows labeled \textit{Sample Generation} and \textit{Sample Verification} indicate the dataset sizes after the \textit{Positive Sample Generation} and \textit{Two-fold Verification} stages, respectively.}}
\label{tb:summary}
\begin{tabular}{l|llllllll|l}
\hline
\multicolumn{1}{c|}{\multirow{2}{*}{\textbf{Stages}}} & \multicolumn{8}{c|}{\textbf{Mutation Rules}}                  & \multicolumn{1}{c}{\multirow{2}{*}{\textbf{Total}}} \\ \cline{2-9}
\multicolumn{1}{c|}{}                                 & FNM   & FPM & CM     & OM     & PM    & FMM & FAM    & Origin & \multicolumn{1}{c}{}                                \\ \hline
Raw Dataset                                           & 0     & 0   & 0      & 0      & 0     & 0   & 0      & 50,000 & 50,000                                              \\
Sample Generation                                     & 1,103 & 668 & 11,717 & 12,560 & 8,061 & 461 & 15,379 & 50,000 & 99,949                                              \\
Sample Verification                                   & 381   & 189 & 5,242  & 6,856  & 2,983 & 176 & 4,140  & 32,160 & 52,127                                              \\ \hline
\end{tabular}
\end{table*}

Due to the numerous costs when invoking the model to achieve sample generation and verification, we randomly selected 50{,}000 commits from the ApacheCM dataset.
As shown in ~\autoref{tb:summary}, the raw dataset initially contains 50{,}000 original negative samples without any mutated instances. 
During the \textit{Sample Generation} stage, diverse mutation rules are applied to create positive samples, resulting in a total of 99{,}949 instances. 
Among the mutation categories, \textit{Feature Adding Mutation} (FAM) produces the largest number of samples, followed by \textit{Operation Mutation} (OM) and \textit{Component Mutation} (CM). 
After applying the \textit{Two-fold Verification} process in the \textit{Sample Verification} stage, low-quality or irrelevant samples are filtered out, reducing the dataset size to 52{,}127 instances. 
This corresponds to an average filtering ratio of approximately 47.84\% compared with the \textit{Sample Generation} stage.
This stage decreases mutated samples across all categories and preserves a large proportion of high-quality data, providing a clean and representative dataset for subsequent model training and evaluation.

\begin{figure*}[h]
\centerline{\includegraphics[width=\linewidth]{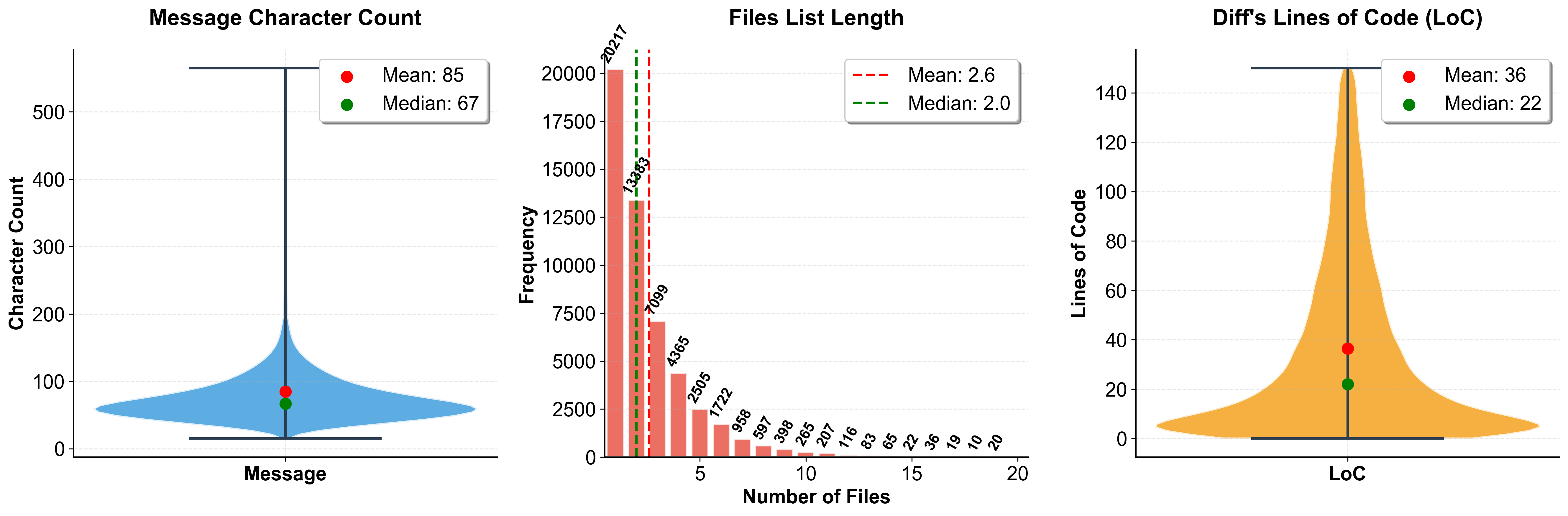}}
	\caption{\textmd{Statistical results of \sys's dataset in terms of \textit{Message character count}, \textit{Diff file count}, and \textit{Diff line of code}.}}
	\label{fg:distribution}
\end{figure*}

We present the statistical results of \sys's dataset in ~\autoref{fg:distribution}.
In summary, the dataset consists of commits from 50 unique GitHub repositories, with an average of 85 characters in the commit message, 2.6 changed files, and 36 diff lines.
Additionally, 75\% of commits consist of fewer than 92 characters in the commit message, 3 changed files, and 64 diff lines.

\subsection{Positive Sample Manual Validation}
\label{sec:manual_validate}

As the positive samples are synthesized by the model, although we introduced \textit{Two-folds Verification} discussed in \autoref{sec:verification} to filter out unsuitable data as much as possible, the dataset still requires manual validation to ensure its validity.
In this section, we propose a statistical analysis to guarantee the achievement of the following goals:

\begin{enumerate}
    \item \textbf{G1: High Correct Rate:} The overall correct rate of the generated positive samples is high.
    \item \textbf{G2: Even Correct Rate Distribution:} The correct rate of positive samples across categories is evenly distributed.
\end{enumerate}

In this analysis, we first randomly selected 100 commits from each type of positive sample.
After manual validation, we found that the correct sample counts for FNM, FPM, CM, OM, PM, FMM, and FAM were 100, 100, 100, 96, 98, 100, and 100, respectively. 
The results indicate that 6 out of the 700 selected samples, which were expected to be inconsistent commits, were actually consistent commits. These discrepancies are limited to operation- and component-related descriptions.
This is because, for operation mutation, the boundaries between some operations are ambiguous. 
For example, when the model generates an "inconsistent" message: \texttt{added/removed code in some place} based on the original message: \texttt{update some code}, these two messages can be considered equivalent by human developers and therefore should not be regarded as inconsistent commits.
For component mutation, however, because a component is an abstract code structure and typically does not appear explicitly in the code diff like file paths or function names, the model sometimes incorrectly identifies the content of the component, leading to a situation where the component has not actually changed before and after mutation.

In the following two paragraphs, we applied the \textit{Wilson confidence interval} to estimate the confidence interval for the overall correctness of the sampling on the full dataset to validate \textbf{G1}, and the \textit{Chi-square goodness-of-fit test} to validate \textbf{G2}.

\paragraph{G1: High Correct Rate}
In total, 694 out of 700 selected samples were verified as correct, resulting in an overall estimated correct ratio of 99.14\%.
The lower bound of the 95\% Wilson confidence interval ($z=1.96$) is 98.14\%, and the upper bound is 99.61\%, indicating that even accounting for sampling bias, the true correct ratio of the dataset remains very high.
Furthermore, the weighted correct ratio, calculated based on the total number of positive samples, is 98.33\%.
These statistics demonstrate that \textit{Two-folds Verification} ensures the overall correctness of the positive samples in the final dataset remains high, achieving \textbf{G1}.

\paragraph{G2: Even Correct Rate Distribution}
Define the null hypothesis ($H_0$): the correct rates for each category are evenly distributed, and the alternative hypothesis ($H_1$): the correct rates for each category are not evenly distributed.
We selected the significance level of $\alpha = 0.05$.
The calculated test statistic is $\chi^2 = 0.1499$, with $p = 0.999934$, which is significantly higher than the corresponding p-threshold for $\alpha = 0.05$.
This result indicates that there is no statistically significant difference in the correct rates for each category. 
The extremely high p-value (0.999934) suggests that the observed differences in correctness can be ruled out as a result of systematic classification bias, thus achieving \textbf{G2}.

\section{Benchmarking for MCI Detection}

The evaluation of \sys aims at answering the following research questions (RQs):

\begin{itemize}
    \item \textbf{RQ1: } How are the basic MCI detection abilities of various LLMs?
    \item \textbf{RQ2: } Can the general LLMs augmentation strategies improve the model's MCI detection ability?
    \item \textbf{RQ3: } Is the MCI detection ability of LLMs evenly distributed under different inconsistency patterns?
\end{itemize}

In this section, we first illustrate the settings specified in our benchmarking, and then we will answer the aforementioned research questions along with the experimental results.

\subsection{Benchmark Settings}

\subsubsection{Evaluation Metrics}

We define the terms used to calculate the evaluation metrics of \sys as follows:

\begin{enumerate}
    \item \textbf{Total Size:} The total size of \sys's dataset, \ie, 52,127.
    \item \textbf{True Positive (TP):} The number of commits whose ground truth is "inconsistent", and the model predicts that it is "inconsistent" as well.
    \item \textbf{False Positive (FP):} The number of commits whose ground truth is "consistent", but the model predicts that it is "inconsistent".
    \item \textbf{True Negative (TN):} The number of commits whose ground truth is "consistent", and the model predicts that it is "consistent" as well.
    \item \textbf{False Negative (FN):} The number of commits whose ground truth is "inconsistent", but the model predicts it is "consistent".
\end{enumerate}

As this benchmark focuses on the models' abilities to distinguish both negative and positive samples, we use \textit{Accuracy}, \textit{Precision}, \textit{Recall}, and \textit{Specificity} to individually present the performance of different models.
We additionally introduce \textit{Matthews Correlation Coefficient (MCC)} as a comprehensive evaluation metric.
These metrics are calculated as follows:

\[
\textit{Accuracy} = \frac{TP + TN}{TP + TN + FP + FN\ (Total\ Size)}
\]

\[
\textit{Precision} = \frac{TP}{TP + FP}
\]

\[
\textit{Recall} = \frac{TP}{TP + FN}
\]

\[
\textit{Specificity} = \frac{TN}{TN + FP}
\]

\[
\textit{MCC} = \frac{TP \times TN - FP \times FN}{\sqrt{(TP + FP)(TP + FN)(TN + FP)(TN + FN)}}
\]

\subsubsection{Benchmarked Models}
In our experiment, we chose \textit{DeepSeek-V3.1}~\footnote{\url{https://huggingface.co/deepseek-ai/DeepSeek-V3.1}} for DeepSeek series model, \textit{Qwen3-30B-A3B-Instruct-2507}~\cite{qwen3technicalreport} for Qwen series model, \textit{gpt-oss-20b}~\footnote{\url{https://huggingface.co/openai/gpt-oss-20b}} for OpenAI's open-weight model, \textit{Llama-3.1-8B-Instruct}~\footnote{\url{https://huggingface.co/meta-llama/Llama-3.1-8B-Instruct}} for Llama series model, \textit{Mistral-Small-3.2-24B-Instruct-2506}\footnote{\url{https://huggingface.co/mistralai/Mistral-Small-3.2-24B-Instruct-2506}} for Mistral AI's series model, and \textit{Kimi-K2-Instruct}~\cite{kimiteam2025kimik2openagentic} for Kimi series model, to evaluate their performances on MCI detection.

\subsubsection{Experimental Settings}
We set the temperature of each model to 0 to obtain stable and reliable outputs. 
For each commit in the dataset, we query the models five times and take the majority decision (\ie, consistent or inconsistent) as the final prediction result for that commit. 
After obtaining the final result, we compute the average number of output tokens that align with the correct answer across all responses, and use this value to represent the token count required for the model to produce the correct answer.
Occasionally, model API requests may fail due to network issues. 
To address this, we allow up to three retries with a 10-second interval between attempts. 
Since all models support long contexts and the lengths of commit messages and code diffs in the dataset are relatively short, no additional context-related configurations are required.

\subsection{RQ1: How are the basic MCI detection abilities of various LLMs?}\label{sec:rq1}

\begin{table*}
\centering
\caption{\textmd{The \textit{Accuracy (Acc.)}, \textit{Precision (Pre.)}, \textit{Recall (Rec.)}, \textit{Specificity (Spec.)}, and \textit{MCC} metrics of models using vanilla prompt templates with no adjacent code lines as context. \textit{Out Tk.} indicates the output token usage of models.}}
\label{tab:vanilla_results}
\resizebox{0.7\linewidth}{!}{
\begin{tabular}{l|l|l|l|l|l|l}
\hline
\multicolumn{1}{c|}{\textbf{Model}} & \multicolumn{1}{c|}{\textbf{Acc.}} & \multicolumn{1}{c|}{\textbf{Pre.}} & \multicolumn{1}{c|}{\textbf{Rec.}} & \multicolumn{1}{c|}{\textbf{Spec.}} & \multicolumn{1}{c|}{\textbf{MCC}} & \multicolumn{1}{c}{\textbf{Out Tk.}} \\ \hline
\textit{DeepSeek-V3.1}              & 83.18                                   & 84.24                                        & 89.46                                     & 73.05                                          & 0.6394                            & 212.48                                       \\
\textit{Qwen3-30B-A3B}              & 74.51                                       & 78.01                                        & 81.73                                     & 62.89                                          & 0.4537                            & 262.50                                       \\
\textit{gpt-oss-20B}                & 85.61                                       & 87.73                                        & 89.14                                     & 79.92                                          & 0.6942                            & 523.44                                       \\
\textit{Llama-3.1-8B}               & 62.82                                       & 64.55                                        & 88.11                                     & 22.08                                          & 0.1358                            & 218.07                                       \\
\textit{Mistral-Small-3.2-24B}      & 75.10                                       & 77.21                                        & 84.61                                     & 59.78                                          & 0.4611                            & 172.04                                       \\
\textit{Kimi-K2-Instruct}                    & 83.59                                       & 89.92                                        & 82.66                                     & 85.08                                          & 0.6646                            & 210.87                                       \\ \hline
\end{tabular}}
\end{table*}

\autoref{tab:vanilla_results} reports the basic performance of different LLMs on the MCI detection task under the vanilla prompt setting, where no adjacent code lines are provided as additional context. 
Overall, \textit{gpt-oss-20B} achieves the best balance among all metrics, reaching the highest accuracy (85.61\%) and MCC (0.6942), indicating a strong overall capability in distinguishing consistent from inconsistent commits. 
\textit{DeepSeek-V3.1} performs comparably well, with slightly lower accuracy (83.18\%) but a higher recall (89.46\%), suggesting a tendency to identify more true inconsistent commits at the cost of lower specificity. 
In contrast, \textit{Kimi-K2} demonstrates the best precision (89.92\%) and a relatively high specificity (85.08\%), showing that it makes fewer false positive predictions, though its recall (82.66\%) is lower than that of \textit{DeepSeek-V3.1}. 

On the other hand, models such as \textit{Qwen3-30B-A3B} and \textit{Mistral-Small-3.2-24B} achieve moderate performance across all metrics, while \textit{Llama-3.1-8B} exhibits the weakest performance, with notably low specificity (22.08\%) and MCC (0.1358), despite maintaining a relatively high recall (88.11\%). 
This indicates that \textit{Llama-3.1-8B} tends to over-predict commits as inconsistent, resulting in many false alarms. 
In terms of efficiency, \textit{Mistral-Small-3.2-24B} produces the lowest average output token usage (172.04), whereas \textit{gpt-oss-20B} incurs the highest cost (523.44).
It presents the trend that when models spend more resources on outputting detailed reasoning, the MCI detection performance will be raised accordingly.
This also suggests a trade-off between performance and inference efficiency. 

\begin{figure*}[th]
	\centerline{\includegraphics[width=0.7\linewidth]{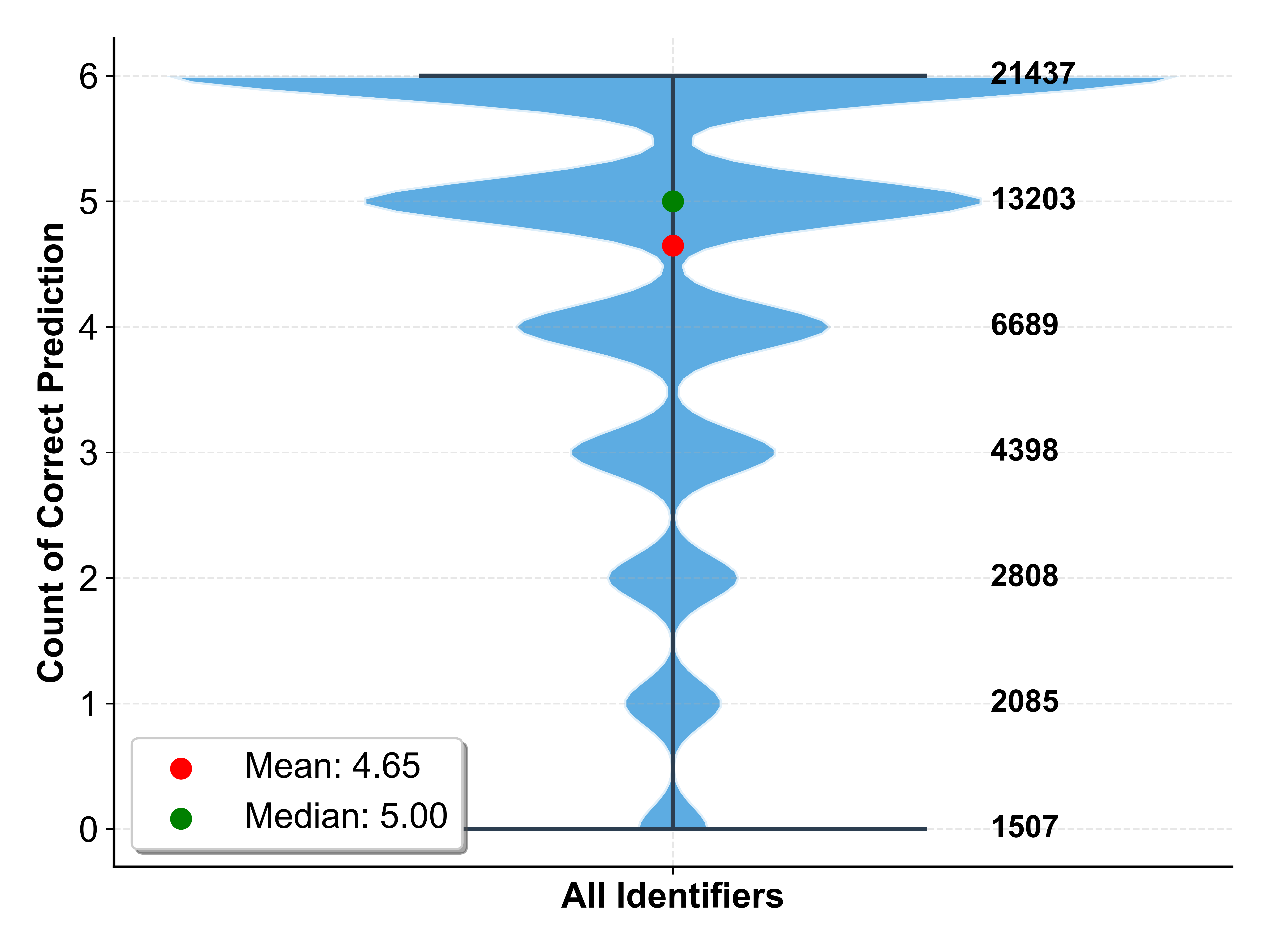}}
	\caption{\textmd{Correct answers' count distribution among the six targeted models.}}
	\label{fg:correct_ans}
\end{figure*}

To obtain a more comprehensive view of the model's performance on MCI detection tasks, we further present the correct answers count distribution in ~\autoref{fg:correct_ans}.
For each commit in \sys's dataset, we record how many models out of the six targeted models gave the correct answer (\ie, consistent or inconsistent).
We found that more than half (66.45\%) of the commits can be correctly classified by most of the models, while still a few (2.89\%) commits cannot be detected properly by any of the models.
By checking out the 1,507 commits, we noticed that 1,272 are positive samples and 235 are negative samples.
This reveals that the models have a worse overall consistency in detecting inconsistent commits than consistent commits.
And since approximately half of the commits in the dataset still prevent the model from producing consistent, correct answers, this further demonstrates that the vanilla template poses significant challenges to achieving overall consistency on the MCI task.

\begin{mdframed}[style=ccl]
\ccl
In our study of \textbf{RQ1}, we observed that all models exhibited strong capabilities in detecting inconsistent commits, achieving an average \textit{Recall} of 85.95\%, along with relatively low false positive rates for positive samples, reflected by an average Precision of 80.28\%.
However, their performance on detecting consistent commits was comparatively weaker, with an average Specificity of only 63.8\%.
Notably, the best-performing model, \textit{gpt-oss-20B}, achieved the highest overall performance but incurred more than twice the output token overhead compared to other models. 
Moreover, while most models successfully identified consistency in over half of the commits, none of them achieved perfect detection, collectively misclassifying 1,272 inconsistent cases and 235 consistent cases. 
\end{mdframed}

\subsection{RQ2: Can the general LLMs augmentation strategies improve the model’s MCI detection ability?}
\label{sec:rq2}

\begin{table}[]
\centering
\caption{\textmd{The \textit{Accuracy (Acc.)}, \textit{Precision (Pre.)}, \textit{Recall (Rec.)}, \textit{Specificity (Spec.)}, and \textit{MCC} metrics of models using extra 20 adjacent code lines as context, few-shots, and chain-of-thought augmentation strategies. \textit{Out Tk.} indicates the output token usage of models. We use arrows with \textcolor{green}{green} background to indicate performance increase compared with the vanilla template without context lines, while arrows with \textcolor{red}{red} background indicate performance decrease.}}
\label{tab:strategies_result}
\resizebox{\linewidth}{!}{
\begin{tabular}{l|c|l|l|l|l|l|l}
\hline
\multicolumn{1}{c|}{\textbf{Model}}             & \textbf{Strategies} & \multicolumn{1}{c|}{\textbf{Acc.}} & \multicolumn{1}{c|}{\textbf{Pre.}} & \multicolumn{1}{c|}{\textbf{Rec.}} & \multicolumn{1}{c|}{\textbf{Spec.}} & \multicolumn{1}{c|}{\textbf{MCC}} & \multicolumn{1}{c}{\textbf{Out Tk.}} \\ \hline
\multirow{3}{*}{\textit{DeepSeek-V3.1}}         & 20 LoC              & 83.09$\textcolor{red}{\downarrow}$                                   & 82.89$\textcolor{red}{\downarrow}$                                        & 91.46$\textcolor{green}{\uparrow}$                                        & 69.59$\textcolor{red}{\downarrow}$                                         & 0.6367$\textcolor{red}{\downarrow}$                          & 210.33$\textcolor{green}{\downarrow}$                                       \\
                                                & few-shots           & 83.41$\textcolor{green}{\uparrow}$                                          & 82.17$\textcolor{red}{\downarrow}$                                      & 93.37$\textcolor{green}{\uparrow}$                                        & 67.36$\textcolor{red}{\downarrow}$                                        & 0.6449$\textcolor{green}{\uparrow}$                               & 179.64$\textcolor{green}{\downarrow}$                                       \\
                                                & CoT                 & 83.34$\textcolor{green}{\uparrow}$                                          & 84.76$\textcolor{green}{\uparrow}$                                           & 89.01$\textcolor{red}{\downarrow}$                                    & 74.22$\textcolor{green}{\uparrow}$                                             & 0.6436$\textcolor{green}{\uparrow}$                               & 421.88$\textcolor{red}{\uparrow}$                                       \\ \hline
\multirow{3}{*}{\textit{Qwen3-30B-A3B}}         & 20 LoC              & 74.52$\textcolor{green}{\uparrow}$                                         & 76.70$\textcolor{red}{\downarrow}$                                        & 84.32$\textcolor{green}{\uparrow}$                                        & 58.74$\textcolor{red}{\downarrow}$                                          & 0.4481$\textcolor{red}{\downarrow}$                         & 267.82$\textcolor{red}{\uparrow}$                                       \\
                                                & few-shots           & 75.12$\textcolor{green}{\uparrow}$                                          & 76.78$\textcolor{red}{\downarrow}$                                      & 85.53$\textcolor{green}{\uparrow}$                                        & 58.35$\textcolor{red}{\downarrow}$                                        & 0.4600$\textcolor{green}{\uparrow}$                               & 279.81$\textcolor{red}{\uparrow}$                                       \\
                                                & CoT                 & 73.69$\textcolor{red}{\downarrow}$                                    & 75.96$\textcolor{red}{\downarrow}$                                       & 83.91$\textcolor{green}{\uparrow}$                                        & 57.22$\textcolor{red}{\downarrow}$                                        & 0.4292$\textcolor{red}{\downarrow}$                          & 684.03$\textcolor{red}{\uparrow}$                                       \\ \hline
\multirow{3}{*}{\textit{gpt-oss-20B}}           & 20 LoC              & 85.93$\textcolor{green}{\uparrow}$                                          & 87.03$\textcolor{red}{\downarrow}$                                      & 90.71$\textcolor{green}{\uparrow}$                                        & 78.23$\textcolor{red}{\downarrow}$                                       & 0.6995$\textcolor{green}{\uparrow}$                               & 561.55$\textcolor{red}{\uparrow}$                                       \\
                                                & few-shots           & 86.29$\textcolor{green}{\uparrow}$                                          & 85.55$\textcolor{red}{\downarrow}$                                      & 93.58$\textcolor{green}{\uparrow}$                                       & 74.54$\textcolor{red}{\downarrow}$                                        & 0.7070$\textcolor{green}{\uparrow}$                               & 490.62$\textcolor{green}{\downarrow}$                                       \\
                                                & CoT                 & 86.11$\textcolor{green}{\uparrow}$                                         & 86.85$\textcolor{red}{\downarrow}$                                       & 91.33$\textcolor{green}{\uparrow}$                                       & 77.72$\textcolor{red}{\downarrow}$                                        & 0.7032$\textcolor{green}{\uparrow}$                              & 676.60$\textcolor{red}{\uparrow}$                                       \\ \hline
\multirow{3}{*}{\textit{Llama-3.1-8B}}          & 20 LoC              & 60.94$\textcolor{red}{\downarrow}$                                      & 64.57$\textcolor{green}{\uparrow}$                                          & 81.28$\textcolor{red}{\downarrow}$                                    & 28.16$\textcolor{green}{\uparrow}$                                            & 0.1103$\textcolor{red}{\downarrow}$                           & 242.99$\textcolor{red}{\uparrow}$                                       \\
                                                & few-shots           & 64.05$\textcolor{green}{\uparrow}$                                         & 67.27$\textcolor{green}{\uparrow}$                                          & 81.25$\textcolor{red}{\downarrow}$                                    & 36.33$\textcolor{green}{\uparrow}$                                            & 0.1962$\textcolor{green}{\uparrow}$                              & 192.18$\textcolor{green}{\downarrow}$                                       \\
                                                & CoT                 & 59.96$\textcolor{red}{\downarrow}$                                      & 74.30$\textcolor{green}{\uparrow}$                                         & 53.66$\textcolor{red}{\downarrow}$                                    & 70.11$\textcolor{green}{\uparrow}$                                            & 0.2324$\textcolor{green}{\uparrow}$                              & 511.42$\textcolor{red}{\uparrow}$                                       \\ \hline
\multirow{3}{*}{\textit{Mistral-Small-3.2-24B}} & 20 LoC              & 73.11$\textcolor{red}{\downarrow}$                                      & 72.73$\textcolor{red}{\downarrow}$                                       & 90.26$\textcolor{green}{\uparrow}$                                       & 45.49$\textcolor{red}{\downarrow}$                                         & 0.4103$\textcolor{red}{\downarrow}$                           & 171.79$\textcolor{green}{\downarrow}$                                       \\
                                                & few-shots           & 75.96$\textcolor{green}{\uparrow}$                                         & 75.47$\textcolor{red}{\downarrow}$                                        & 90.43$\textcolor{green}{\uparrow}$                                        & 52.66$\textcolor{red}{\downarrow}$                                          & 0.4771$\textcolor{green}{\uparrow}$                              & 142.13$\textcolor{green}{\downarrow}$                                       \\
                                                & CoT                 & 75.50$\textcolor{green}{\uparrow}$                                         & 74.82$\textcolor{red}{\downarrow}$                                       & 90.87$\textcolor{green}{\uparrow}$                                       & 50.74$\textcolor{red}{\downarrow}$                                         & 0.4667$\textcolor{green}{\uparrow}$                              & 383.24$\textcolor{red}{\uparrow}$                                       \\ \hline
\multirow{3}{*}{\textit{Kimi-K2-Instruct}}               & 20 LoC              & 84.33$\textcolor{green}{\uparrow}$                                         & 89.05$\textcolor{red}{\downarrow}$                                       & 85.06$\textcolor{green}{\uparrow}$                                      & 83.15$\textcolor{red}{\downarrow}$                                         & 0.6740$\textcolor{green}{\uparrow}$                              & 215.97$\textcolor{red}{\uparrow}$                                       \\
                                                & few-shots           & 86.16$\textcolor{green}{\uparrow}$                                         & 86.86$\textcolor{red}{\downarrow}$                                        & 91.39$\textcolor{green}{\uparrow}$                                       & 77.73$\textcolor{red}{\downarrow}$                                          & 0.7041$\textcolor{green}{\uparrow}$                              & 177.63$\textcolor{green}{\downarrow}$                                       \\
                                                & CoT                 & 83.26$\textcolor{red}{\downarrow}$                                       & 94.30$\textcolor{green}{\uparrow}$                                          & 77.56$\textcolor{green}{\uparrow}$                                     & 92.45$\textcolor{green}{\uparrow}$                                            & 0.6807$\textcolor{green}{\uparrow}$                              & 652.11$\textcolor{red}{\uparrow}$                                       \\ \hline
\end{tabular}}
\end{table}

The results in ~\autoref{tab:strategies_result} reveal several noteworthy patterns regarding the effectiveness of different augmentation strategies across models.
First, the effect of providing adjacent code lines (20 LoC) is model-dependent.
For large models such as \textit{gpt-oss-20B}, 20 LoC raises accuracy from the vanilla baseline to 85.93\% and MCC to 0.6995, representing improvements over the base configuration. 
Similarly, \textit{Kimi-K2-Instruct} achieves 84.33\% accuracy and 0.6740 MCC. 
In contrast, smaller models show degradation: 
\textit{Llama-3.1-8B} drops to 60.94\% accuracy and 0.1103 MCC, indicating that extra context introduces noise rather than signal for under-parameterized models.
Second, few-shot prompting is the most consistently beneficial strategy.
For example, \textit{DeepSeek-V3.1} increases accuracy to 83.41\% and MCC to 0.6449, while simultaneously reducing output tokens from 210.33 (20 LoC) to 179.64. 
\textit{Mistral-Small-3.2-24B} shows a similar pattern: 
Accuracy rises from 73.11\% (20 LoC) to 75.96\%, and MCC improves from 0.4103 to 0.4771, with output tokens dropping from 171.79 to 142.13. 
These results suggest that explicit demonstrations provide stable gains while lowering computational costs.
Third, CoT prompting reveals a precision–recall trade-off.
For \textit{Kimi-K2-Instruct}, CoT achieves the highest precision (94.30\%) and specificity (92.45\%) across all settings, but recall falls sharply to 77.56\%, compared to 91.39\% under few-shot. 
This indicates that CoT enforces stricter decision boundaries, favoring correctness over coverage. 
A similar effect is observed in \textit{Llama-3.1-8B}, where precision rises to 74.30\%, yet recall collapses to 53.66\%, suggesting overfitting to narrow reasoning chains.

Finally, output token usage highlights the efficiency–performance trade-off.
Few-shot prompting consistently reduces token usage (e.g., \textit{DeepSeek-V3.1:} 179.64 vs. 210.33, \textit{Kimi-K2-Instruct:} 177.63 vs. 215.97) while maintaining or improving accuracy.
In contrast, CoT substantially inflates token consumption.
For example, \textit{Qwen3-30B-A3B} reaches 684.03 tokens, more than double the 20 LoC strategy (267.82 tokens), without corresponding performance benefits.

We offer the following rationale for the aforementioned observations.
First, few-shot exemplars serve as direct induction of the decision pattern. By showing the model how to map from commit message + diff to a JSON judgment plus concise reasoning, the model can more confidently follow the same mapping with minimal “reasoning” overhead. 
This template-based guidance constrains the model’s generation space, thereby reducing verbose or extraneous output. 
In contrast, neither 20 LoC nor CoT provides an explicit exemplar of how to compress reasoning while making the consistency judgment.
Second, the 20 LoC strategy burdens the attention mechanism with noisy or irrelevant information, which can dilute the signal-to-noise ratio. The LLM must expend capacity to filter contextual noise and determine relevance, which often triggers the model to generate extra justification text or restatements, thus increasing token consumption without reliably improving accuracy.
Third, CoT encourages verbose, stepwise reasoning that is more appropriate for deep multi-hop reasoning tasks. 
In our experiment, forcing full reasoning chains often leads the model to unpack every diff block and re-explain trivial parts, inflating the output. 
Additionally, CoT explanations are not always faithful to the internal decision process, and may be biased or post-hoc rationalizations~\cite{turpin2023languagemodelsdontsay}.
Meanwhile, the few-shot prompt already implicitly encodes a compact reasoning style, making explicit CoT redundant and even counterproductive.

\begin{figure*}[th]
	\centerline{\includegraphics[width=\linewidth]{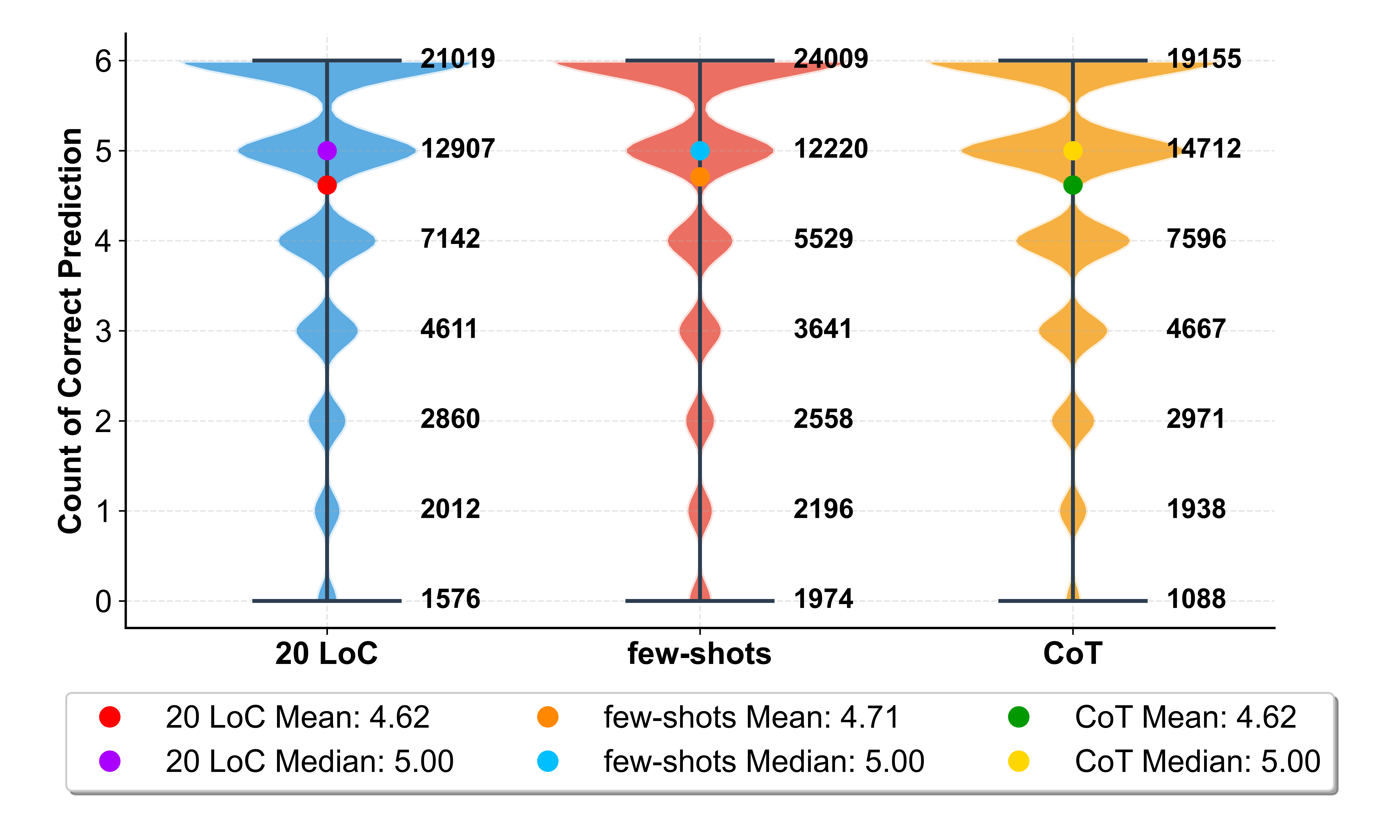}}
	\caption{\textmd{Correct answers' count distribution among the six targeted models with the three augmentation strategies.}}
	\label{fg:correct_ans_strate}
\end{figure*}

Similarly to ~\autoref{sec:rq1}, we present ~\autoref{fg:correct_ans_strate} to illustrate the correct answers count distribution across targeted models with 20 LoC, few-shots, and CoT augmentation strategies, respectively.
Compared to the vanilla models (shown in ~\autoref{fg:correct_ans}), we observed that, with the exception of few-shot prompting, the average correct answer counts of the 20 LoC and CoT strategies declined slightly. 
This suggests that 20 LoC and CoT are more likely to induce divergent judgments across models on the same commit, thereby reducing overall consistency in MCI detection. 
In contrast, the few-shot strategy maintains stronger consistency, enabling a greater number of models to converge on the same detection outcome.
Surprisingly, although the few-shot strategy achieves the highest overall consistency, it also produces the largest number of commits for which no model makes a correct prediction: 
1,974 cases, comprising 1,737 inconsistent commits and 237 consistent commits. 
By contrast, the 20 LoC and CoT strategies result in fewer universally misdetected commits, with 1,576 cases (1,354 inconsistent and 222 consistent) and 1,088 cases (518 inconsistent and 570 consistent), respectively.
This is due to the corresponding commit's message and code diff relationships diverging significantly from the patterns represented in the few-shot exemplars. 
Because few-shot prompting relies heavily on in-context example induction, the model may fail entirely when confronted with a commit whose structure or rationale lies outside the exemplar manifold. 
In other words, the demonstration bias inherent in few-shot learning can lead to poor generalization on edge cases.
Second, few-shot prompting can introduce a kind of template bias, the language model may overcommit to the inference patterns implicit in the exemplars, even when they mismatch the actual commit structure. 
For some commits, rather than fall back to more exploratory reasoning, the model may rigidly follow exemplar-driven heuristics and hence systematically err across all models.
To prove the above explanation about the abnormal correct answer distribution, we stepped into ~\autoref{sec:rq3} to check these strategies' performances on various inconsistent commit types.

\begin{mdframed}[style=ccl]
\ccl
In our study of \textbf{RQ2}, it demonstrates that augmentation strategies exert heterogeneous effects on model performance in MCI detection. 
While providing adjacent code lines benefits larger models, it introduces noise for smaller ones.
Few-shot prompting emerges as the most reliable approach in MCI detection, consistently improving accuracy and MCC while also reducing output token consumption. 
However, its high overall consistency comes at the cost of producing the largest set of universally misclassified commits, revealing a trade-off between stability and coverage. 
By contrast, CoT prompting enforces stricter precision–recall boundaries, improving precision and specificity at the expense of recall, but substantially inflating token usage. 
Taken together, these findings highlight that no single strategy uniformly dominates across dimensions of accuracy, efficiency, and robustness, but rather each entails distinct trade-offs that must be carefully balanced depending on application requirements.
\end{mdframed}

\subsection{RQ3: Is the MCI detection ability of LLMs evenly distributed under different inconsistency patterns?}\label{sec:rq3}

\begin{table}[]
\centering
\caption{\textmd{The detection ratio of various augmentation strategies and inconsistent commit types. Since these are all positive samples (\ie, inconsistent commits), we use the ratio of the number of correctly detected inconsistent commits to the total number of commits in this category as the detection ratio.  We use a darker cell background to indicate better performance (\ie. higher detection ratio).}}
\label{tab:grouped_detect_ratio}
\resizebox{\linewidth}{!}{
\begin{tabular}{l|l|l|l|l|l|l|l|l}
\hline
\multicolumn{1}{c|}{\textbf{Model}}                                      & \multicolumn{1}{c|}{\textbf{Strategies}} & \multicolumn{1}{c|}{\textbf{CM}} & \multicolumn{1}{c|}{\textbf{FAM}} & \multicolumn{1}{c|}{\textbf{FMM}} & \multicolumn{1}{c|}{\textbf{FPM}} & \multicolumn{1}{c|}{\textbf{FNM}} & \multicolumn{1}{c|}{\textbf{OM}} & \multicolumn{1}{c}{\textbf{PM}} \\ \hline
\cellcolor[HTML]{FFCCC9}                                                 & \cellcolor[HTML]{FFCCC9}CoT              & \cellcolor[HTML]{FF5858}76.19    & \cellcolor[HTML]{FF0E00}87.71     & \cellcolor[HTML]{FF8686}75.57     & \cellcolor[HTML]{FF3838}78.31     & \cellcolor[HTML]{FFD2D2}61.68     & \cellcolor[HTML]{FFB2B2}74.90    & \cellcolor[HTML]{FFE4E4}51.76   \\ \cline{2-9} 
\cellcolor[HTML]{FFCCC9}                                                 & \cellcolor[HTML]{FFCCC9}few-shots        & \cellcolor[HTML]{FFB2B2}64.88    & \cellcolor[HTML]{FF0E00}82.83     & \cellcolor[HTML]{FF8686}65.34     & \cellcolor[HTML]{FF5858}70.37     & \cellcolor[HTML]{FFD2D2}49.08     & \cellcolor[HTML]{FF3838}70.84    & \cellcolor[HTML]{FFE4E4}44.49   \\ \cline{2-9} 
\cellcolor[HTML]{FFCCC9}                                                 & \cellcolor[HTML]{FFCCC9}vanilla          & \cellcolor[HTML]{FF5858}76.84    & \cellcolor[HTML]{FF0E00}84.30     & \cellcolor[HTML]{FFB2B2}73.30     & \cellcolor[HTML]{FF3838}78.84     & \cellcolor[HTML]{FFD2D2}65.88     & \cellcolor[HTML]{FF8686}73.63    & \cellcolor[HTML]{FFE4E4}49.98   \\ \cline{2-9} 
\multirow{-4}{*}{\cellcolor[HTML]{FFCCC9}\textit{DeepSeek-V3.1}}         & \cellcolor[HTML]{FFCCC9}20 LoC           & \cellcolor[HTML]{FF5858}75.14    & \cellcolor[HTML]{FF0E00}79.35     & \cellcolor[HTML]{FFD2D2}65.34     & \cellcolor[HTML]{FF3838}78.84     & \cellcolor[HTML]{FFB2B2}67.72     & \cellcolor[HTML]{FF8686}70.49    & \cellcolor[HTML]{FFE4E4}44.15   \\ \hline
\cellcolor[HTML]{FFCE93}                                                 & \cellcolor[HTML]{FFCE93}CoT              & \cellcolor[HTML]{FFB052}86.36    & \cellcolor[HTML]{FF8B00}95.43     & \cellcolor[HTML]{FFCE94}78.98     & \cellcolor[HTML]{FF9F2B}87.30     & \cellcolor[HTML]{FFC176}81.89     & \cellcolor[HTML]{FFDDB4}72.02    & \cellcolor[HTML]{FFEAD0}49.82   \\ \cline{2-9} 
\cellcolor[HTML]{FFCE93}                                                 & \cellcolor[HTML]{FFCE93}few-shots        & \cellcolor[HTML]{FFB052}80.96    & \cellcolor[HTML]{FF8B00}95.12     & \cellcolor[HTML]{FFDDB4}70.45     & \cellcolor[HTML]{FF9F2B}88.36     & \cellcolor[HTML]{FFC176}73.75     & \cellcolor[HTML]{FFCE94}70.54    & \cellcolor[HTML]{FFEAD0}43.38   \\ \cline{2-9} 
\cellcolor[HTML]{FFCE93}                                                 & \cellcolor[HTML]{FFCE93}vanilla          & \cellcolor[HTML]{FFB052}88.42    & \cellcolor[HTML]{FF8B00}95.80     & \cellcolor[HTML]{FFCE94}81.25     & \cellcolor[HTML]{FF9F2B}90.48     & \cellcolor[HTML]{FFC176}83.20     & \cellcolor[HTML]{FFDDB4}75.38    & \cellcolor[HTML]{FFEAD0}52.23   \\ \cline{2-9} 
\multirow{-4}{*}{\cellcolor[HTML]{FFCE93}\textit{gpt-oss-20B}}           & \cellcolor[HTML]{FFCE93}20 LoC           & \cellcolor[HTML]{FFB052}87.79    & \cellcolor[HTML]{FF8B00}92.42     & \cellcolor[HTML]{FFCE94}81.82     & \cellcolor[HTML]{FF9F2B}89.42     & \cellcolor[HTML]{FFC176}86.88     & \cellcolor[HTML]{FFDDB4}73.95    & \cellcolor[HTML]{FFEAD0}49.55   \\ \hline
\cellcolor[HTML]{FFFC9E}                                                 & \cellcolor[HTML]{FFFC9E}CoT              & \cellcolor[HTML]{FFFF2B}94.14    & \cellcolor[HTML]{FFFF00}98.60     & \cellcolor[HTML]{FDFF8E}89.77     & \cellcolor[HTML]{FFFF49}93.65     & \cellcolor[HTML]{FFFFC3}86.88     & \cellcolor[HTML]{FDFF7D}92.27    & \cellcolor[HTML]{FFFEDF}82.17   \\ \cline{2-9} 
\cellcolor[HTML]{FFFC9E}                                                 & \cellcolor[HTML]{FFFC9E}few-shots        & \cellcolor[HTML]{FFFF49}81.88    & \cellcolor[HTML]{FFFF00}90.43     & \cellcolor[HTML]{FDFF8E}73.86     & \cellcolor[HTML]{FFFF2B}84.13     & \cellcolor[HTML]{FFFFC3}69.55     & \cellcolor[HTML]{FDFF7D}76.75    & \cellcolor[HTML]{FFFEDF}55.95   \\ \cline{2-9} 
\cellcolor[HTML]{FFFC9E}                                                 & \cellcolor[HTML]{FFFC9E}vanilla          & \cellcolor[HTML]{FFFF2B}91.51    & \cellcolor[HTML]{FFFF00}94.28     & \cellcolor[HTML]{FDFF7D}84.66     & \cellcolor[HTML]{FFFF49}91.01     & \cellcolor[HTML]{FFFFC3}82.15     & \cellcolor[HTML]{FDFF8E}82.31    & \cellcolor[HTML]{FFFEDF}67.38   \\ \cline{2-9} 
\multirow{-4}{*}{\cellcolor[HTML]{FFFC9E}\textit{Kimi-K2-Instruct}}      & \cellcolor[HTML]{FFFC9E}20 LoC           & \cellcolor[HTML]{FFFF49}91.07    & \cellcolor[HTML]{FFFF2B}91.11     & \cellcolor[HTML]{FDFF8E}82.95     & \cellcolor[HTML]{FFFF00}91.53     & \cellcolor[HTML]{FDFF7D}83.46     & \cellcolor[HTML]{FFFFC3}80.89    & \cellcolor[HTML]{FFFEDF}62.79   \\ \hline
\cellcolor[HTML]{CEFFCE}                                                 & \cellcolor[HTML]{CEFFCE}CoT              & \cellcolor[HTML]{4FFA4D}76.99    & \cellcolor[HTML]{C2FEC2}62.42     & \cellcolor[HTML]{9EFF9D}65.91     & \cellcolor[HTML]{03FF00}78.84     & \cellcolor[HTML]{79FF77}68.50     & \cellcolor[HTML]{32FF30}77.87    & \cellcolor[HTML]{E2FEE2}50.72   \\ \cline{2-9} 
\cellcolor[HTML]{CEFFCE}                                                 & \cellcolor[HTML]{CEFFCE}few-shots        & \cellcolor[HTML]{9EFF9D}32.11    & \cellcolor[HTML]{32FF30}39.59     & \cellcolor[HTML]{79FF77}36.93     & \cellcolor[HTML]{4FFA4D}38.10     & \cellcolor[HTML]{E2FEE2}24.93     & \cellcolor[HTML]{03FF00}41.15    & \cellcolor[HTML]{C2FEC2}29.47   \\ \cline{2-9} 
\cellcolor[HTML]{CEFFCE}                                                 & \cellcolor[HTML]{CEFFCE}vanilla          & \cellcolor[HTML]{79FF77}23.33    & \cellcolor[HTML]{C2FEC2}17.44     & \cellcolor[HTML]{4FFA4D}26.14     & \cellcolor[HTML]{03FF00}31.75     & \cellcolor[HTML]{9EFF9D}18.11     & \cellcolor[HTML]{32FF30}26.23    & \cellcolor[HTML]{E2FEE2}16.43   \\ \cline{2-9} 
\multirow{-4}{*}{\cellcolor[HTML]{CEFFCE}\textit{Llama-3.1-8B}}          & \cellcolor[HTML]{CEFFCE}20 LoC           & \cellcolor[HTML]{4FFA4D}31.02    & \cellcolor[HTML]{E2FEE2}20.97     & \cellcolor[HTML]{9EFF9D}28.41     & \cellcolor[HTML]{03FF00}46.56     & \cellcolor[HTML]{79FF77}29.40     & \cellcolor[HTML]{32FF30}32.86    & \cellcolor[HTML]{C2FEC2}20.99   \\ \hline
\cellcolor[HTML]{ECF4FF}                                                 & \cellcolor[HTML]{ECF4FF}CoT              & \cellcolor[HTML]{2782FF}62.51    & \cellcolor[HTML]{ECF4FF}34.66     & \cellcolor[HTML]{9BC5FF}49.43     & \cellcolor[HTML]{006BFF}69.84     & \cellcolor[HTML]{72ADFF}53.28     & \cellcolor[HTML]{549CFF}57.04    & \cellcolor[HTML]{B9D7FF}36.41   \\ \cline{2-9} 
\cellcolor[HTML]{ECF4FF}                                                 & \cellcolor[HTML]{ECF4FF}few-shots        & \cellcolor[HTML]{9BC5FF}50.90    & \cellcolor[HTML]{549CFF}56.52     & \cellcolor[HTML]{72ADFF}52.84     & \cellcolor[HTML]{006BFF}59.79     & \cellcolor[HTML]{B9D7FF}44.36     & \cellcolor[HTML]{2782FF}58.55    & \cellcolor[HTML]{ECF4FF}37.45   \\ \cline{2-9} 
\cellcolor[HTML]{ECF4FF}                                                 & \cellcolor[HTML]{ECF4FF}vanilla          & \cellcolor[HTML]{2782FF}73.62    & \cellcolor[HTML]{B9D7FF}44.88     & \cellcolor[HTML]{9BC5FF}64.20     & \cellcolor[HTML]{006BFF}79.89     & \cellcolor[HTML]{549CFF}66.67     & \cellcolor[HTML]{72ADFF}65.13    & \cellcolor[HTML]{ECF4FF}41.47   \\ \cline{2-9} 
\multirow{-4}{*}{\cellcolor[HTML]{ECF4FF}\textit{Mistral-Small-3.2-24B}} & \cellcolor[HTML]{ECF4FF}20 LoC           & \cellcolor[HTML]{2782FF}64.99    & \cellcolor[HTML]{ECF4FF}22.29     & \cellcolor[HTML]{72ADFF}51.70     & \cellcolor[HTML]{006BFF}74.60     & \cellcolor[HTML]{549CFF}54.59     & \cellcolor[HTML]{9BC5FF}50.32    & \cellcolor[HTML]{B9D7FF}28.90   \\ \hline
\cellcolor[HTML]{CBCEFB}                                                 & \cellcolor[HTML]{CBCEFB}CoT              & \cellcolor[HTML]{B774FF}54.24    & \cellcolor[HTML]{9F48FC}55.65     & \cellcolor[HTML]{EAD5FF}45.45     & \cellcolor[HTML]{8B1DFE}61.38     & \cellcolor[HTML]{CA97FF}48.03     & \cellcolor[HTML]{7C00FF}65.59    & \cellcolor[HTML]{D8B5FC}47.00   \\ \cline{2-9} 
\cellcolor[HTML]{CBCEFB}                                                 & \cellcolor[HTML]{CBCEFB}few-shots        & \cellcolor[HTML]{B774FF}49.35    & \cellcolor[HTML]{8B1DFE}55.36     & \cellcolor[HTML]{D8B5FC}48.86     & \cellcolor[HTML]{9F48FC}55.03     & \cellcolor[HTML]{EAD5FF}47.24     & \cellcolor[HTML]{7C00FF}72.05    & \cellcolor[HTML]{CA97FF}48.98   \\ \cline{2-9} 
\cellcolor[HTML]{CBCEFB}                                                 & \cellcolor[HTML]{CBCEFB}vanilla          & \cellcolor[HTML]{9F48FC}59.98    & \cellcolor[HTML]{B774FF}59.66     & \cellcolor[HTML]{CA97FF}54.55     & \cellcolor[HTML]{8B1DFE}64.55     & \cellcolor[HTML]{EAD5FF}52.49     & \cellcolor[HTML]{7C00FF}71.56    & \cellcolor[HTML]{D8B5FC}54.31   \\ \cline{2-9} 
\multirow{-4}{*}{\cellcolor[HTML]{CBCEFB}\textit{Qwen3-30B-A3B}}         & \cellcolor[HTML]{CBCEFB}20 LoC           & \cellcolor[HTML]{9F48FC}55.82    & \cellcolor[HTML]{B774FF}52.03     & \cellcolor[HTML]{EAD5FF}44.32     & \cellcolor[HTML]{8B1DFE}64.55     & \cellcolor[HTML]{CA97FF}51.18     & \cellcolor[HTML]{7C00FF}69.09    & \cellcolor[HTML]{D8B5FC}50.82   \\ \hline
\end{tabular}}
\end{table}

\begin{table}[]
\centering
\caption{\textmd{The average output token usage of various augmentation strategies and inconsistent commit types. We use a darker cell background to indicate better performance (\ie. less token usage).}}
\label{tab:grouped_tokens}
\resizebox{\linewidth}{!}{
\begin{tabular}{l|l|l|l|l|l|l|l|l}
\hline
\multicolumn{1}{c|}{\textbf{Model}}                                      & \multicolumn{1}{c|}{\textbf{Strategies}} & \multicolumn{1}{c|}{\textbf{CM}} & \multicolumn{1}{c|}{\textbf{FAM}} & \multicolumn{1}{c|}{\textbf{FMM}} & \multicolumn{1}{c|}{\textbf{FPM}} & \multicolumn{1}{c|}{\textbf{FNM}} & \multicolumn{1}{c|}{\textbf{OM}} & \multicolumn{1}{c}{\textbf{PM}} \\ \hline
\cellcolor[HTML]{FFCCC9}                                                 & \cellcolor[HTML]{FFCCC9}CoT              & \cellcolor[HTML]{FF5858}441.98   & \cellcolor[HTML]{FFD2D2}452.68    & \cellcolor[HTML]{FFE4E4}469.82    & \cellcolor[HTML]{FF0000}425.00    & \cellcolor[HTML]{FFB2B2}447.38    & \cellcolor[HTML]{FF3838}432.05   & \cellcolor[HTML]{FF8686}442.80  \\ \cline{2-9} 
\cellcolor[HTML]{FFCCC9}                                                 & \cellcolor[HTML]{FFCCC9}few-shots        & \cellcolor[HTML]{FFB2B2}199.94   & \cellcolor[HTML]{FFE4E4}215.34    & \cellcolor[HTML]{FFD2D2}214.52    & \cellcolor[HTML]{FF0000}193.46    & \cellcolor[HTML]{FF8686}199.39    & \cellcolor[HTML]{FF5858}196.27   & \cellcolor[HTML]{FF3838}193.71  \\ \cline{2-9} 
\cellcolor[HTML]{FFCCC9}                                                 & \cellcolor[HTML]{FFCCC9}vanilla          & \cellcolor[HTML]{FFB2B2}236.61   & \cellcolor[HTML]{FFD2D2}247.39    & \cellcolor[HTML]{FFE4E4}256.30    & \cellcolor[HTML]{FF0000}225.85    & \cellcolor[HTML]{FF8686}234.56    & \cellcolor[HTML]{FF5858}232.87   & \cellcolor[HTML]{FF3838}229.29  \\ \cline{2-9} 
\multirow{-4}{*}{\cellcolor[HTML]{FFCCC9}\textit{DeepSeek-V3.1}}         & \cellcolor[HTML]{FFCCC9}20 LoC           & \cellcolor[HTML]{FF8686}236.56   & \cellcolor[HTML]{FFD2D2}246.64    & \cellcolor[HTML]{FFE4E4}253.14    & \cellcolor[HTML]{FF3838}227.04    & \cellcolor[HTML]{FFB2B2}236.62    & \cellcolor[HTML]{FF5858}232.02   & \cellcolor[HTML]{FF0000}224.07  \\ \hline
\cellcolor[HTML]{FFCE93}                                                 & \cellcolor[HTML]{FFCE93}CoT              & \cellcolor[HTML]{FF9F2B}685.43   & \cellcolor[HTML]{FFCE94}739.30    & \cellcolor[HTML]{FFB052}692.99    & \cellcolor[HTML]{FF8B00}577.92    & \cellcolor[HTML]{FFC176}710.40    & \cellcolor[HTML]{FFDDB4}812.92   & \cellcolor[HTML]{FFEAD0}892.31  \\ \cline{2-9} 
\cellcolor[HTML]{FFCE93}                                                 & \cellcolor[HTML]{FFCE93}few-shots        & \cellcolor[HTML]{FF9F2B}478.82   & \cellcolor[HTML]{FFC176}557.20    & \cellcolor[HTML]{FFB052}500.80    & \cellcolor[HTML]{FF8B00}455.15    & \cellcolor[HTML]{FFEAD0}863.56    & \cellcolor[HTML]{FFCE94}605.68   & \cellcolor[HTML]{FFDDB4}632.40  \\ \cline{2-9} 
\cellcolor[HTML]{FFCE93}                                                 & \cellcolor[HTML]{FFCE93}vanilla          & \cellcolor[HTML]{FF9F2B}498.49   & \cellcolor[HTML]{FFC176}601.82    & \cellcolor[HTML]{FFB052}558.01    & \cellcolor[HTML]{FF8B00}494.29    & \cellcolor[HTML]{FFCE94}635.22    & \cellcolor[HTML]{FFDDB4}637.33   & \cellcolor[HTML]{FFEAD0}697.05  \\ \cline{2-9} 
\multirow{-4}{*}{\cellcolor[HTML]{FFCE93}\textit{gpt-oss-20B}}           & \cellcolor[HTML]{FFCE93}20 LoC           & \cellcolor[HTML]{FF8B00}526.66   & \cellcolor[HTML]{FFDDB4}709.69    & \cellcolor[HTML]{FFCE94}635.99    & \cellcolor[HTML]{FF9F2B}556.63    & \cellcolor[HTML]{FFB052}590.72    & \cellcolor[HTML]{FFC176}628.18   & \cellcolor[HTML]{FFEAD0}727.41  \\ \hline
\cellcolor[HTML]{FFFC9E}                                                 & \cellcolor[HTML]{FFFC9E}CoT              & \cellcolor[HTML]{FFFF2B}674.24   & \cellcolor[HTML]{FDFF8E}682.80    & \cellcolor[HTML]{FFFEDF}723.20    & \cellcolor[HTML]{FFFF00}637.21    & \cellcolor[HTML]{FDFF7D}681.29    & \cellcolor[HTML]{FFFF49}676.41   & \cellcolor[HTML]{FFFFC3}707.25  \\ \cline{2-9} 
\cellcolor[HTML]{FFFC9E}                                                 & \cellcolor[HTML]{FFFC9E}few-shots        & \cellcolor[HTML]{FDFF7D}200.54   & \cellcolor[HTML]{FFFFC3}217.06    & \cellcolor[HTML]{FFFEDF}217.53    & \cellcolor[HTML]{FFFF00}194.95    & \cellcolor[HTML]{FFFF49}198.49    & \cellcolor[HTML]{FDFF8E}201.16   & \cellcolor[HTML]{FFFF2B}197.91  \\ \cline{2-9} 
\cellcolor[HTML]{FFFC9E}                                                 & \cellcolor[HTML]{FFFC9E}vanilla          & \cellcolor[HTML]{FFFF2B}234.98   & \cellcolor[HTML]{FFFFC3}240.33    & \cellcolor[HTML]{FFFEDF}259.76    & \cellcolor[HTML]{FFFF00}225.06    & \cellcolor[HTML]{FFFF49}235.50    & \cellcolor[HTML]{FDFF7D}235.63   & \cellcolor[HTML]{FDFF8E}237.76  \\ \cline{2-9} 
\multirow{-4}{*}{\cellcolor[HTML]{FFFC9E}\textit{Kimi-K2-Instruct}}      & \cellcolor[HTML]{FFFC9E}20 LoC           & \cellcolor[HTML]{FFFF49}239.32   & \cellcolor[HTML]{FFFFC3}246.15    & \cellcolor[HTML]{FFFEDF}261.11    & \cellcolor[HTML]{FFFF00}228.78    & \cellcolor[HTML]{FFFF2B}237.11    & \cellcolor[HTML]{FDFF7D}240.63   & \cellcolor[HTML]{FDFF8E}242.03  \\ \hline
\cellcolor[HTML]{E6FEE5}                                                 & \cellcolor[HTML]{E6FEE5}CoT              & \cellcolor[HTML]{C2FEC2}533.25   & \cellcolor[HTML]{E2FEE2}552.73    & \cellcolor[HTML]{79FF77}522.99    & \cellcolor[HTML]{32FF30}519.97    & \cellcolor[HTML]{9EFF9D}527.24    & \cellcolor[HTML]{4FFA4D}519.97   & \cellcolor[HTML]{03FF00}518.99  \\ \cline{2-9} 
\cellcolor[HTML]{E6FEE5}                                                 & \cellcolor[HTML]{E6FEE5}few-shots        & \cellcolor[HTML]{79FF77}195.01   & \cellcolor[HTML]{E2FEE2}235.95    & \cellcolor[HTML]{C2FEC2}202.74    & \cellcolor[HTML]{03FF00}181.49    & \cellcolor[HTML]{32FF30}189.23    & \cellcolor[HTML]{4FFA4D}189.64   & \cellcolor[HTML]{9EFF9D}197.77  \\ \cline{2-9} 
\cellcolor[HTML]{E6FEE5}                                                 & \cellcolor[HTML]{E6FEE5}vanilla          & \cellcolor[HTML]{79FF77}220.39   & \cellcolor[HTML]{E2FEE2}237.98    & \cellcolor[HTML]{4FFA4D}219.17    & \cellcolor[HTML]{03FF00}205.04    & \cellcolor[HTML]{9EFF9D}222.29    & \cellcolor[HTML]{32FF30}215.40   & \cellcolor[HTML]{C2FEC2}227.06  \\ \cline{2-9} 
\multirow{-4}{*}{\cellcolor[HTML]{E6FEE5}\textit{Llama-3.1-8B}}          & \cellcolor[HTML]{E6FEE5}20 LoC           & \cellcolor[HTML]{9EFF9D}247.10   & \cellcolor[HTML]{E2FEE2}260.54    & \cellcolor[HTML]{32FF30}233.51    & \cellcolor[HTML]{03FF00}223.48    & \cellcolor[HTML]{79FF77}243.15    & \cellcolor[HTML]{4FFA4D}241.93   & \cellcolor[HTML]{C2FEC2}247.43  \\ \hline
\cellcolor[HTML]{ECF4FF}                                                 & \cellcolor[HTML]{ECF4FF}CoT              & \cellcolor[HTML]{72ADFF}406.46   & \cellcolor[HTML]{B9D7FF}413.38    & \cellcolor[HTML]{ECF4FF}425.09    & \cellcolor[HTML]{006BFF}383.19    & \cellcolor[HTML]{9BC5FF}406.71    & \cellcolor[HTML]{2782FF}390.59   & \cellcolor[HTML]{549CFF}393.24  \\ \cline{2-9} 
\cellcolor[HTML]{ECF4FF}                                                 & \cellcolor[HTML]{ECF4FF}few-shots        & \cellcolor[HTML]{B9D7FF}152.17   & \cellcolor[HTML]{ECF4FF}172.67    & \cellcolor[HTML]{72ADFF}158.40    & \cellcolor[HTML]{006BFF}148.92    & \cellcolor[HTML]{9BC5FF}153.59    & \cellcolor[HTML]{2782FF}151.36   & \cellcolor[HTML]{549CFF}151.84  \\ \cline{2-9} 
\cellcolor[HTML]{ECF4FF}                                                 & \cellcolor[HTML]{ECF4FF}vanilla          & \cellcolor[HTML]{72ADFF}185.98   & \cellcolor[HTML]{ECF4FF}189.45    & \cellcolor[HTML]{B9D7FF}187.39    & \cellcolor[HTML]{549CFF}182.15    & \cellcolor[HTML]{9BC5FF}186.15    & \cellcolor[HTML]{2782FF}178.38   & \cellcolor[HTML]{006BFF}175.99  \\ \cline{2-9} 
\multirow{-4}{*}{\cellcolor[HTML]{ECF4FF}\textit{Mistral-Small-3.2-24B}} & \cellcolor[HTML]{ECF4FF}20 LoC           & \cellcolor[HTML]{9BC5FF}185.36   & \cellcolor[HTML]{72ADFF}185.00    & \cellcolor[HTML]{ECF4FF}186.30    & \cellcolor[HTML]{549CFF}183.30    & \cellcolor[HTML]{B9D7FF}185.52    & \cellcolor[HTML]{2782FF}175.83   & \cellcolor[HTML]{006BFF}174.30  \\ \hline
\cellcolor[HTML]{DFC1FF}                                                 & \cellcolor[HTML]{DFC1FF}CoT              & \cellcolor[HTML]{B774FF}685.91   & \cellcolor[HTML]{7C00FF}647.63    & \cellcolor[HTML]{CA97FF}687.45    & \cellcolor[HTML]{8B1DFE}656.14    & \cellcolor[HTML]{EAD5FF}1375.40   & \cellcolor[HTML]{D8B5FC}751.71   & \cellcolor[HTML]{9F48FC}664.65  \\ \cline{2-9} 
\cellcolor[HTML]{DFC1FF}                                                 & \cellcolor[HTML]{DFC1FF}few-shots        & \cellcolor[HTML]{B774FF}296.07   & \cellcolor[HTML]{D8B5FC}336.69    & \cellcolor[HTML]{CA97FF}307.95    & \cellcolor[HTML]{EAD5FF}1654.93   & \cellcolor[HTML]{9F48FC}290.87    & \cellcolor[HTML]{7C00FF}285.93   & \cellcolor[HTML]{8B1DFE}290.22  \\ \cline{2-9} 
\cellcolor[HTML]{DFC1FF}                                                 & \cellcolor[HTML]{DFC1FF}vanilla          & \cellcolor[HTML]{9F48FC}279.56   & \cellcolor[HTML]{EAD5FF}315.15    & \cellcolor[HTML]{D8B5FC}290.86    & \cellcolor[HTML]{7C00FF}267.90    & \cellcolor[HTML]{B774FF}280.39    & \cellcolor[HTML]{8B1DFE}273.07   & \cellcolor[HTML]{CA97FF}281.43  \\ \cline{2-9} 
\multirow{-4}{*}{\cellcolor[HTML]{DFC1FF}\textit{Qwen3-30B-A3B}}         & \cellcolor[HTML]{DFC1FF}20 LoC           & \cellcolor[HTML]{B774FF}286.38   & \cellcolor[HTML]{EAD5FF}328.85    & \cellcolor[HTML]{9F48FC}282.99    & \cellcolor[HTML]{7C00FF}271.77    & \cellcolor[HTML]{CA97FF}287.34    & \cellcolor[HTML]{8B1DFE}279.94   & \cellcolor[HTML]{D8B5FC}287.37  \\ \hline
\end{tabular}}
\end{table}

Simply exploring the evaluation metrics across the positive and negative samples is not enough to grasp the detailed MCI detection ability distribution.
This is because different types of inconsistent commit types potentially pose various complications for MCI detection.
Besides, the uneven distribution of \sys's dataset (see ~\autoref{tb:summary}) may introduce bias to the benchmarking results.
To this end, we split the results of inconsistent commit types and present ~\autoref{tab:grouped_detect_ratio}.
We use the ratio of the number of correctly detected inconsistent commits to the total number of commits in this category as the detection ratio.
Additionally, we illustrate the average output token usage of models with different augmentation strategies in detecting the inconsistent commit types in ~\autoref{tab:grouped_tokens}.

\autoref{tab:grouped_detect_ratio} reports the detection ratios of several LLMs for different inconsistent commit types (CM, FAM, FMM, FPM, FNM, OM, PM) under four augmentation strategies: CoT, few-shots, vanilla, and 20 LoC. 
From the table, it is evident that model capability strongly affects performance. 
\textit{Kimi-K2-Instruct} consistently achieves the highest detection ratios across almost all types, reaching over 90\% for CM, FAM, and FPM with CoT and vanilla strategies. 
\textit{gpt-oss-20B} also demonstrates strong performance, particularly for CM and FAM, where CoT and vanilla strategies yield detection ratios above 85\%. 
\textit{DeepSeek-V3.1} shows moderate performance, generally in the 65–85\% range, while \textit{Llama-3.1-8B} and \textit{Mistral-Small-3.2-24B} exhibit lower detection ratios, with some types, such as FMM and PM, dropping below 40\% under limited-context strategies. 
\textit{Qwen3-30B-A3B} achieves intermediate performance, often below gpt-oss-20B and Kimi-K2-Instruct but above \textit{Llama-3.1-8B}. 
Augmentation strategy also plays a key role. 
CoT generally outperforms few-shot, vanilla, and 20 LoC, suggesting that reasoning-guided prompts help models detect subtle inconsistencies. 
Few-shot often yields lower detection ratios than CoT and vanilla, particularly for \textit{Llama-3.1-8B} and \textit{Mistral-Small-3.2-24B}, indicating sensitivity to limited exemplars. 
The 20 LoC strategy shows mixed results: for high-capacity models like \textit{Kimi-K2-Instruct} and \textit{gpt-oss-20B}, performance remains strong, while smaller models suffer substantial drops, especially for FNM and PM. 
Vanilla strategies usually produce moderate detection ratios, providing a baseline for comparison. 
The impact of inconsistent types is also pronounced. CM, FAM, and FPM are generally easier to detect, with many models achieving ratios above 75–80\%, whereas FNM and PM remain challenging for most models except the top-performing ones. 
OM exhibits intermediate difficulty, with detection ratios varying substantially between models and strategies. 

\autoref{tab:grouped_tokens} presents the average output token usage of various LLMs for different inconsistent commit types under four augmentation strategies.
From the table, it is clear that model architecture and strategy substantially influence token usage. For models like \textit{Kimi-K2-Instruct} and \textit{gpt-oss-20B}, CoT generally results in higher token consumption, often exceeding 600–700 tokens for CM, FAM, and FMM. 
However, these models achieve lower token usage under few-shot, with values dropping to around 200–300 tokens, indicating that exemplar-based prompting substantially reduces output length while maintaining performance. 
Vanilla and 20 LoC strategies exhibit intermediate token usage, typically between CoT and few-shot, highlighting a trade-off between reasoning depth and output conciseness.
In contrast, models such as \textit{Llama-3.1-8B}, \textit{Mistral-Small-3.2-24B}, and \textit{DeepSeek-V3.1} show a similar pattern but with overall lower token usage. 
For example, \textit{Llama-3.1-8B} under CoT consumes approximately 520–550 tokens for most types, whereas few-shot reduces this to around 180–200 tokens. 
Notably, certain inconsistent types, such as OM and PM, require more tokens under CoT, reaching up to 890 tokens in \textit{gpt-oss-20B} or 1,375 tokens in \textit{Qwen3-30B-A3B}, reflecting the complexity of reasoning needed for these types.
Across all models, few-shot consistently leads to the lowest token usage, particularly for simpler inconsistent types such as CM, FMM, and FPM. 
The 20 LoC strategy shows moderate efficiency, suggesting that limiting context size can reduce token output but may require careful balance to avoid performance degradation.

From the above analysis, we noticed that CM, FPM, and OM commits often manifest as direct contrasts between the commit message and the code diff. 
They are typically surface-level or structural conflicts that can be resolved via simple matching or logical comparison. 
For instance, commit messages frequently include verbs such as add, remove, update, fix, and delete, along with explicit targets; 
Meanwhile, the code diff clearly presents “+” (added) lines, “–” (deleted) lines, and contextual changes. 
If a message states “add X” but the diff shows that X is deleted, the inconsistency is immediately obvious.
At the same time, because the contradictory elements in these cases can usually be matched, compared, and located directly within the message and diff, the model does not need to construct long intermediate reasoning steps or perform elaborate inference. 
This allows it to make judgments more efficiently, resulting in relatively low usage of output tokens.
In contrast, the detection rate for purpose inconsistencies (PM) is the lowest, while the corresponding token usage is substantially higher. 
This phenomenon arises because identifying the underlying intent of a commit requires strong code comprehension capabilities. 
Many commit purposes are high-level or repository-specific, making them difficult, or even impossible, to infer solely from the code diff. 
Moreover, this type of inconsistency lies at a deep semantic level, where detection demands a longer reasoning path. 
The model often needs to interpret contextual background and analyze potential deviations in intent, resulting in more verbose reasoning and, consequently, higher token consumption.

Regarding the anomalous performance of the few-shot strategy, which exhibits the highest count of commits with no correct detections as shown in~\autoref{sec:rq2}, we observe that it underperforms in detecting CM, OM, and PM inconsistencies compared to the 20 LoC, vanilla, and CoT strategies. 
Upon examining the distribution of inconsistent commit types in~\autoref{tb:summary}, we further find that these three categories constitute a substantial portion of the positive samples. 
Consequently, the overall detection performance of the few-shot strategy may be suppressed because of the unevenly distributed dataset, leading to consistent misclassification of these commits across multiple models.

\begin{mdframed}[style=ccl]
\ccl
In our study of \textbf{RQ3}, it shows that the models' MCI detection performances are not evenly distributed under different inconsistency patterns.
Specifically, CM, FPM, and OM inconsistencies exhibit higher detection ratios and lower token consumption, as their message–code contradictions are explicit and can be resolved through direct textual or structural comparison. 
In contrast, PM inconsistencies demand deeper semantic understanding and contextual reasoning, resulting in lower detection accuracy and significantly higher token usage. 
Furthermore, although the few-shot strategy demonstrates strong performance in general, its effectiveness is constrained by the data imbalance across inconsistency types, particularly for CM, OM, and PM commits, which dominate the dataset. 
This imbalance leads to suppressed overall detection performance and higher rates of uniform incorrect detection results across models. 
These observations collectively suggest that improving detection of high-level, intent-driven inconsistencies requires not only richer contextual grounding but also more balanced training and evaluation data.

\end{mdframed}

\subsection{Case Studies}

To better illustrate the general poor performance on purpose inconsistency, and the increased count of no-correct detection results from few-shot, we picked the following two case studies to show how models with different strategies gave the correct/incorrect detection results.

\subsubsection{Case Study 1 – Inability to Be Augmented}
\autoref{lst:casestudy1} presents an inconsistent commit case from \textit{Apache Doris}~\footnote{\url{https://github.com/apache/doris.git}}.
In this commit, none of the strategies can help with generating the correct detection results.

The code diff shows the removal of a debug line (i.e., line 14). 
Since the code change is simple and the commit message describes code change does align with the actual code change, all augmentation strategies, including the vanilla template, detect that the message aligns with the actual code change, as shown in ~\autoref{lst:casestudy1_output}. 
However, the commit message also states that the purpose of this change is to "improve monitoring," which is misleading because merely removing the debug line does not enhance monitoring performance. 
Upon inspecting the original commit, we found that the actual reason for this change was to fix a bug where the debug point could cause a page issue~\footnote{\url{https://github.com/apache/doris/commit/95fc7a28d28789aa95e2125755a1b649ea8b145a}}. 
Therefore, this commit should be considered inconsistent. 
Unfortunately, the model cannot produce any accurate detection results, even with the augmentation strategies.
This case study illustrates that the three augmentation strategies lack the capability of improving the model's reasoning regarding the purpose or intent of the commit.

\begin{lstlisting}[label={lst:casestudy1}, language=diff, caption={\textmd{Inconsistent commit case, which no augmentation strategies can be leveraged to provide the correct detection result.}}]
@@ **Commit message:**
@@ feat(load): enhance bytes_appended accuracy for 
@@ improved monitoring by removing debug injection point.

diff --git a/be/src/io/fs/file_writer.h b/be/src/io/fs/file_writer.h
index bb3235e7d27e4d..6996f2e0c983ac 100644
--- a/be/src/io/fs/file_writer.h
+++ b/be/src/io/fs/file_writer.h
@@ -64,10 +64,7 @@ class FileWriter {
 
     const Path& path() const { return _path; }
 
-    size_t bytes_appended() const {
-        DBUG_EXECUTE_IF("FileWriter.bytes_appended.zero_bytes_appended", { return 0; });
-        return _bytes_appended;
-    }
+    size_t bytes_appended() const { return _bytes_appended; }
 
     std::shared_ptr<FileSystem> fs() const { return _fs; }

\end{lstlisting}

\begin{lstlisting}[label={lst:casestudy1_output}, language=diff, caption={\textmd{Corresponding outputs of gpt-oss-20b with augmentation strategies.}}]
@@ Vanilla prompt
# Is consistent: 
True
# Reasoning: 
The commit message states that the change enhances the accuracy of `bytes_appended` for improved monitoring by removing a debug injection point. 
The diff shows exactly that: the `DBUG_EXECUTE_IF` macro that could return 0 has been removed, leaving a straightforward return of `_bytes_appended`. 
Thus the message accurately describes the code change. 

@@ 20 LoC
# Is consistent: 
True
# Reasoning:
The commit message states that the change enhances the accuracy of `bytes_appended` by removing a debug injection point. 
The diff shows exactly that: the `DBUG_EXECUTE_IF` macro that could return 0 was removed, leaving a straightforward return of `_bytes_appended`.
No other code changes are present, so the message accurately describes the modification. 

@@ few-shots
# Is consistent: 
True
# Reasoning:
The commit message states that the change enhances the accuracy of `bytes_appended()` for better monitoring by removing a debug injection point. 
The diff shows exactly that: the `DBUG_EXECUTE_IF` macro that could return 0 has been removed, leaving a straightforward return of `_bytes_appended`. 
No other files or functions are modified, and the scope of the change (removing a debug stub) matches the message. 
Therefore the message accurately describes the code changes

@@ CoT
# Is consistent: 
True
# Reasoning:
Step 1: The commit message claims to improve the accuracy of the `bytes_appended` metric by removing a debug injection point, which should lead to better monitoring. 
Step 2: The diff shows a single change in `be/src/io/fs/file_writer.h`: the `bytes_appended()` method is simplified to return `_bytes_appended` directly, removing a `DBUG_EXECUTE_IF` block that could return 0. This directly eliminates the debug injection that could distort the metric. 
Step 3: The message matches the code change exactly: the debug injection is removed, and the method now returns the true value, improving accuracy. No other files or functions are touched, and the scope (minor change to a single method) aligns with the message. 
Step 4: No discrepancies are found; the message accurately describes the change. 
Step 5: Therefore, the commit message is consistent with the code diff.

\end{lstlisting}

\subsubsection{Case Study 2 - Few-Shot Strategy Performance Degradation}
\autoref{lst:casestudy2} shows a inconsistent commit case from \textit{Apache Incubator-weex}~\footnote{\url{https://github.com/apache/incubator-weex}}.
In this commit, none of the models using the few-shot augmentation strategy produced correct detection results. 
In contrast, the vanilla \textit{Kimi-K2-Instruct} model successfully identified the inconsistency and provided meaningful reasoning.

The code diff shows that a parent object is assigned to \texttt{domObject}, which is of type \texttt{WXDomObject}. 
However, the commit message states that the parent is added to \texttt{renderObject}. 
This discrepancy indicates that the commit message does not accurately describe the actual code modification. 
For few-shot strategies, all models interpreted the inconsistency between \texttt{WXDomObject} and \texttt{renderObject} merely as a terminology difference within the codebase (\eg, lines 14–17 of ~\autoref{lst:casestudy2_output}), and therefore did not mark this as an inconsistency between the code diff and the message. 
Conversely, the vanilla \textit{Kimi-K2-Instruct} model, as shown in lines 6–9 of ~\autoref{lst:casestudy2_output}, correctly concluded that the message and the code diff were inconsistent, as the object to which the parent was added was not of type \texttt{renderObject}. 
Additionally, we also observed that the vanilla model incorrectly reported another inconsistency regarding the change scope. 
This false positive arose from the absence of contextual code lines. 
Specifically, the model inferred that the modification was fixing a missing line rather than adding new functionality “when it is created.” 
This misunderstanding occurred because, aside from the 20 LoC variant, the prompt only includes the diff lines without surrounding context. 
In this commit, the code diff provided to the vanilla prompt contained only lines 4–8 and 12–13 (with line numbers adjusted accordingly), preventing the model from recognizing that \texttt{domObject} was created from a JSON object at line 10.

\begin{lstlisting}[label={lst:casestudy2}, language=diff, caption={\textmd{Inconsistent commit case, which can be detected by vanilla Kimi-K2-Instruct model, but cannot be detected by that model with few-shots augmentation.}}]
@@ **Commit message:**
@@ add parent to renderObject, when it is created

diff --git a/android/sdk/src/main/java/com/taobao/weex/dom/WXDomObject.java b/android/sdk/src/main/java/com/taobao/weex/dom/WXDomObject.java
index 983b500d34..dab3d2ece3 100644
--- a/android/sdk/src/main/java/com/taobao/weex/dom/WXDomObject.java
+++ b/android/sdk/src/main/java/com/taobao/weex/dom/WXDomObject.java
@@ -663,7 +663,7 @@ public String dumpDomTree() {
       }
       domObject.parseFromJson(json);
       domObject.mDomContext = wxsdkInstance;
- 
+      domObject.parent = parentDomObject;
       Object children = json.get(CHILDREN);
       if (children != null && children instanceof JSONArray) {
         JSONArray childrenArray = (JSONArray) children;


\end{lstlisting}

\begin{lstlisting}[label={lst:casestudy2_output}, language=diff, caption={\textmd{Corresponding outputs of Kimi-K2-Instruct with no and few-shots augmentation strategies.}}]

@@ Kimi-K2-Instruct Vanilla
# Is consistent:
False
# Reasoning:
The commit message claims to 'add parent to renderObject, when it is created', but the actual code change is setting the parent on a 'domObject' (specifically a WXDomObject), not a renderObject. 
The message mentions 'renderObject', which is a different concept from 'domObject' in the Weex framework. 
Additionally, the change appears to be fixing an assignment that was previously missing or removed (indicated by the '-' showing a blank line being replaced with the assignment), rather than adding new functionality 'when it is created'. 
The scope and terminology in the commit message do not accurately reflect the actual change made.

@@ Kimi-K2-Instruct few-shot
# Is consistent:
True
# Reasoning:
The commit message states 'add parent to renderObject, when it is created' and the code change shows the addition of a line that sets 'domObject.parent = parentDomObject'. 
While the message mentions 'renderObject' and the actual change is in 'WXDomObject' (which appears to be a DOM object rather than a render object), this is likely a terminology issue within the codebase where DOM objects and render objects are closely related.
The core action described in the message - adding a parent reference when the object is created - is accurately reflected in the code change.

\end{lstlisting}
\section{Threat to Validity}

\subsection{Uneven Distribution of Synthesized Dataset}

As shown in ~\autoref{tb:summary}, while we enhanced the quality of positive samples using robust LLMs and two-fold validation, and demonstrated dataset validity through manual verification and statistical methods in ~\autoref{sec:manual_validate}, this does not ensure a uniform distribution of inconsistency types within the positive samples. 
This is because, during the \textit{Positive Sample Generation} phase, we filter out mutation rules that are inconsistent with the original commit message to ensure reasonable synthetic samples. 
For instance, if the original commit message does not mention a file path, FPM is excluded as a mutation rule. 
Arbitrarily introducing a file path could make the inconsistency more noticeable and unrealistic, thus inflating the detection accuracy of inconsistency by the targeted models/strategies. 
This filtering process results in certain mutation rules, such as FMM and FPM, being applied to only a small subset of commit messages, leading to an uneven distribution of inconsistency types.
This uneven distribution could bias the overall performance evaluation, favoring the inconsistency types that dominate the dataset. 
To address this issue, we report results by inconsistency type in ~\autoref{sec:rq3} to assess the MCI detection capabilities of the models and strategies under different types of inconsistency. 
Nevertheless, the differing data sizes of these types may still introduce some bias in the results.

\subsection{More Dimensions of Prompts}

In the \sys benchmarking, we evaluated four model configurations for MCI detection: the vanilla model, contextual adjacent code lines, few-shot, and CoT strategies. 
While the current results have yielded several valuable preliminary findings, such as the overall superiority of the few-shot strategy in detecting inconsistencies and the increased output token usage caused by CoT reasoning, there remains significant potential for more fine-grained experiments to fully uncover the limitations of these augmentation strategies in MCI detection.
For the contextual adjacent code lines strategy, experiments can be extended by varying the number of provided context lines (\eg, 10/20/50/100 LoC) to examine how different context lengths influence detection performance.
For the few-shot strategy, future work could vary the number of examples or construct specialized examples based on the anomalous cases identified in our case studies to investigate whether targeted demonstrations improve detection accuracy.
For the CoT strategy, alternative reasoning paths could be explored to determine whether different reasoning formulations lead to more reliable MCI detection outcomes.
Therefore, the experimental results presented in the current \sys benchmark reflect only the findings under the specific experimental configurations adopted in this benchmark and may not generalize to other MCI detection settings or model configurations.
\section{Related Works}

\subsection{Message–Code Inconsistency Detection}

A growing body of work addresses the detection of inconsistencies between source code and its natural-language annotations, such as document-code inconsistency and comment-code inconsistency.
Early techniques relied on hand-crafted rules or program analysis to flag obvious mismatches (\eg, checking nullness and value ranges in comments ~\cite{ratol2017detecting}). 
More recent approaches leverage deep learning or large language models. For instance, Panthaplackel \etal ~\cite{panthaplackel2021deep} formulate comment–code consistency as a natural-language inference task, training a neural model to detect just-in-time mismatches between code changes and method comments. 
Steiner and Zhang ~\cite{steiner2022code} similarly train Transformer-based encoders (BERT/Longformer) to classify code–comment pairs as consistent or not, outperforming earlier feature-based baselines. 
Hybrid static–ML methods have also been proposed: Xie \etal ~\cite{xu2024detecting} introduce MPDetector, which combines symbolic execution of Python libraries with LLM-driven extraction of API constraints to catch inconsistencies in multi-parameter documentation. 
Likewise, Hao \etal ~\cite{hao2023smartcoco} present SmartCoco, a domain-specific approach that uses constraint propagation for consistency checking in smart contract comments. 
Recent advances increasingly exploit pretrained code LLMs: 
For example, Dau \etal ~\cite{dau2023docchecker} present DocChecker, a fine-tuned model that identifies and even corrects code–comment mismatches, achieving new state-of-the-art accuracy on standard benchmarks. 
Empirical studies (\eg, Wen \etal ~\cite{wen2019large}) have quantified how often such inconsistencies arise and cataloged the types of comment updates fixed by developers. 
In summary, the literature has moved from simple syntactic checks toward learning-based and hybrid techniques (including LLMs) to capture the semantic misalignment between code and its comments or documentation.

\subsection{Commit Message Generation Methods}

Automatic generation of commit messages has likewise seen rapid progress. Early systems used template or rule-based approaches (\eg, Changescribe ~\cite{linares2015changescribe}) or treated the task as machine translation. 
Neural encoder–decoder models were first applied to diffs: Jiang \etal ~\cite{jiang2017automatically} and Loyola \etal ~\cite{loyola2017neural} use sequence-to-sequence RNNs to “translate” code changes into natural-language descriptions.
Subsequent work enhanced these models with attention and copy mechanisms.
For example, Xu \etal ~\cite{xu2019commit} introduce CoDiSum, a bidirectional-GRU model that better captures code semantics.
These studies show that neural approaches can outperform basic NMT baselines ~\cite{liu2018neural,tao2021evaluation}.

Beyond purely generative models, retrieval-based and hybrid methods have been proposed. 
Recently, Shi \etal ~\cite{shi2022race} introduce RACE, a retrieval-augmented generation approach: 
RACE retrieves exemplar commits with similar diffs and uses them to guide the neural decoder, yielding more accurate and less repetitive messages. 
Similarly, He \etal ~\cite{he2023come} propose COME, which encodes fine-grained modification embeddings of code edits and combines translation and retrieval results to improve message quality.

The latest trend is to leverage large pretrained models. 
Zhang \etal ~\cite{zhang2024using} perform a preliminary study showing that open- and closed-source LLMs (\eg,  LLaMA, ChatGPT) often generate more human-like messages than prior models, even if traditional metrics (BLEU/ROUGE) do not always reflect this. 
Xue \etal ~\cite{xue2024automated} conduct a comprehensive empirical comparison and find that state-of-the-art LLMs outperform conventional commit-generation models on both automatic and human-evaluated criteria.
They further propose a retrieval-based in-context learning method (ERICommiter) that boosts LLM performance by selecting exemplar commits as prompts. 
These advances demonstrate that neural and LLM-based models now dominate the state of the art in commit message generation, moving beyond early NMT approaches.

\subsection{Commit Quality Benchmarking and Empirical Studies}

Assessing commit message quality has become an active concern. 
Much work in commit generation uses overlap-based metrics (BLEU, ROUGE, METEOR) for evaluation ~\cite{tao2021evaluation}, but these often do not correlate well with human judgments. 
Schatzle \etal ~\cite{schall2024commitbench} address this by constructing CommitBench, a filtered, multilingual dataset of high-quality commits and introducing a specialized metric (C-GOOD) to better measure adequacy of generated messages. 
They show that training on this curated dataset yields more informative commit descriptions. Other researchers have advocated for semantic evaluation: for example, Zhang \etal ~\cite{zhang2024using} experiment with LLM-based evaluators as scalable proxies for human raters.

Empirical studies have also examined real commit histories. 
Li and Ahmed ~\cite{commitstudy1} perform an empirical study on open-source repositories and find that low-quality or inconsistent commit messages are about 1.5 times more likely to precede bug-introducing changes, and that overall commit message quality tends to degrade over the lifetime of a project. 
Complementing this, Corral \etal ~\cite{farago2023full} survey industry and academic guidelines and develop a machine-learning-based checker to automatically assess commit messages for semantic quality. 
Together, these benchmarking efforts and empirical analyses underscore the impact of message quality on software maintenance and motivate improved metrics and tools for evaluating commit messages.
\section{Conclusion}

In this paper, we propose \sys, a rich and type-aware benchmarking towards LLMs' commit message and code change inconsistency detection ability.
To obtain the benchmarking dataset with both positive and negative samples of commits, we construct \sys's dataset based on the ApacheCM dataset.
For each existing consistent commit in the original dataset, we cherry-pick the most suitable mutation rules from the seven pre-defined and the most common inconsistent types.
https://cn.overleaf.com/project%
To ensure the validity of the synthesized inconsistent commits, we then applied a two-fold verification atop the power of LLM to automatically filter out inadequate commits, and proceeded with randomly-sampled manual verification in order to prove the validity of the dataset in a statistical way.
We evaluated six state-of-the-art open-source LLMs with four MCI detection templates: vanilla model, few-shot prompting, contextual adjacent code lines, and CoT.
The results show that the basic inconsistent commit detection performance of \textit{gpt-oss-20b} prevails over other models', but in turn, consumes almost 2 times the output token usage.
Moreover, the benchmarking also reveals that augmentation strategy effects are heterogeneous: contextual adjacent code lines help larger models but can add noise for smaller ones; few-shot generally boosts accuracy and reduces token usage, yet increases universally incorrect cases; CoT improves precision and specificity at the cost of recall and token usage.
Ultimately, type-wise analysis reveals higher detectability for component, file-path, and operation inconsistencies, and markedly lower accuracy with higher token cost for intent-level “purpose” inconsistencies. 

\section{Acknowledgments}
This work was supported by Ant Group Research Intern Program.


\bibliographystyle{colm2024_conference}
\bibliography{custom}

\clearpage
\appendix
\section{Benchmarking Templates}\label{apd:templates}

\begin{lstlisting}[label={lst:vanilla_template}, language=diff, caption={\textmd{Vanilla prompting template.}}]
@@ System prompt
You are an expert code reviewer tasked with evaluating the consistency between commit messages and their corresponding code changes.

Your job is to analyze whether a commit message accurately describes the actual code changes made in the diff.

Consider the following aspects:
1. Does the commit message describe the actual changes made?
2. Are the mentioned components/files/functions actually modified?
3. Is the scope of changes (major/minor) consistent with the message?
4. Are any important changes missing from the commit message?
5. Does the commit message contain any false or misleading information?

Respond with a JSON object containing:
 - "consistent": true/false,
 - "reasoning": "detailed explanation of your analysis",
 - "issues": ["list of specific inconsistencies found, if any"]

Be thorough in your analysis and provide clear reasoning for your decision.

@@ User prompt
Please analyze the consistency between this commit message and the corresponding code diff:

**Commit Message:**
{commit_message}

**Code Diff:**
{code_diff}

Evaluate whether the commit message accurately describes the code changes and respond with the requested JSON format.

\end{lstlisting}
\begin{lstlisting}[label={lst:fewshot_template}, language=diff, caption={\textmd{Few-shot prompting template.}}]
@@ System prompt
You are an expert code reviewer tasked with evaluating the consistency between commit messages and their corresponding code changes.

Your job is to analyze whether a commit message accurately describes the actual code changes made in the diff.

Consider the following aspects:
1. Does the commit message describe the actual changes made?
2. Are the mentioned components/files/functions actually modified?
3. Is the scope of changes (major/minor) consistent with the message?
4. Are any important changes missing from the commit message?
5. Does the commit message contain any false or misleading information?

Here are two examples to guide your analysis:

**Example 1 (Consistent):**
Commit Message: "Fix memory leak in buffer allocation"
Code Diff:
```
 void allocate_buffer() {{
 -    char* buf = malloc(1024);
 +    char* buf = malloc(1024);
 +    if (!buf) return;
      // process buffer
 +    free(buf);
 }}
```
Analysis: {{
  "consistent": true,
  "reasoning": "The commit message accurately describes the change. A memory leak was indeed fixed by adding proper error checking and freeing the allocated buffer.",
  "issues": []
}}

**Example 2 (Inconsistent):**
Commit Message: "Add new sorting algorithm implementation"
Code Diff:
```
 int compare_strings(const char* a, const char* b) {{
 -    return strcmp(a, b);
 +    return strcasecmp(a, b);
 }}
```
Analysis: {{
  "consistent": false,
  "reasoning": "The commit message claims to add a new sorting algorithm, but the actual change only modifies a string comparison function to be case-insensitive. No sorting algorithm was added.",
  "issues": ["Commit message mentions adding sorting algorithm but only string comparison was changed", "Scope mismatch: minor change vs claimed major addition"]
}}

Respond with a JSON object containing:
 - "consistent": true/false,
 - "reasoning": "detailed explanation of your analysis",
 - "issues": ["list of specific inconsistencies found, if any"]

Be thorough in your analysis and provide clear reasoning for your decision.

@@ User prompt
Please analyze the consistency between this commit message and the corresponding code diff:

**Commit Message:**
{commit_message}

**Code Diff:**
{code_diff}

Evaluate whether the commit message accurately describes the code changes and respond with the requested JSON format.

\end{lstlisting}
\begin{lstlisting}[label={lst:cot_template}, language=diff, caption={\textmd{CoT prompting template.}}]
@@ System prompt
You are an expert code reviewer tasked with evaluating the consistency between commit messages and their corresponding code changes.

Your job is to analyze whether a commit message accurately describes the actual code changes made in the diff using a step-by-step Chain of Thought approach.

Please follow this structured reasoning process:

**Step 1: Understand the Commit Message**
 - What is the main purpose/goal described in the commit message?
 - What specific changes or fixes does it claim to make?
 - What components, files, or functionality does it mention?

**Step 2: Analyze the Code Diff**
 - What files are actually modified?
 - What specific code changes are made (additions, deletions, modifications)?
 - What functionality is actually being changed or implemented?

**Step 3: Compare Message vs Reality**
 - Do the claimed changes in the message match the actual code changes?
 - Are the mentioned components/files actually modified?
 - Is the scope of changes (major/minor) consistent with the message?
 - Are there any important changes in the code that are not mentioned in the message?
 - Are there any claims in the message that are not supported by the code changes?

**Step 4: Identify Inconsistencies (if any)**
 - List specific discrepancies between the message and code
 - Note any misleading or false information in the commit message
 - Identify missing information that should have been mentioned

**Step 5: Make Final Decision**
 - Based on the analysis above, determine if the message is consistent with the code changes
 - Assign a confidence level (0.0-1.0) based on the clarity and strength of the evidence
 - Provide a clear reasoning for the decision

Consider the following aspects in your analysis:
1. Does the commit message describe the actual changes made?
2. Are the mentioned components/files/functions actually modified?
3. Is the scope of changes (major/minor) consistent with the message?
4. Are any important changes missing from the commit message?
5. Does the commit message contain any false or misleading information?

Respond with a JSON object containing:
 - "consistent": true/false,
 - "reasoning": "detailed step-by-step explanation following the Chain of Thought process above",
 - "issues": ["list of specific inconsistencies found, if any"]

Be thorough in your step-by-step analysis and provide clear reasoning for your decision."""


@@ User prompt
Please analyze the consistency between this commit message and the corresponding code diff using the Chain of Thought approach:

**Commit Message:**
{commit_message}

**Code Diff:**
{code_diff}

Follow the 5-step Chain of Thought process outlined in the system prompt to evaluate whether the commit message accurately describes the code changes, and respond with the requested JSON format.

\end{lstlisting}

\appendix
\clearpage

\end{document}